\documentclass[10pt,journal,twoside,final]{IEEEtran}

\IEEEoverridecommandlockouts

\makeatletter
\newcommand{\biggg}[1]{{\hbox{$\left#1\vbox to 20.5pt{}\right.\n@space$}}}
\newcommand{\Biggg}[1]{{\hbox{$\left#1\vbox to 23.5pt{}\right.\n@space$}}}
\newcommand{\bigggg}[1]{{\hbox{$\left#1\vbox to 26.5pt{}\right.\n@space$}}}
\newcommand{\Bigggg}[1]{{\hbox{$\left#1\vbox to 29.5pt{}\right.\n@space$}}}
\newcommand{\biggggg}[1]{{\hbox{$\left#1\vbox to 32.5pt{}\right.\n@space$}}}
\newcommand{\Biggggg}[1]{{\hbox{$\left#1\vbox to 35.5pt{}\right.\n@space$}}}
\newcommand{\bigggggg}[1]{{\hbox{$\left#1\vbox to 38.5pt{}\right.\n@space$}}}
\newcommand{\Bigggggg}[1]{{\hbox{$\left#1\vbox to 41.5pt{}\right.\n@space$}}}
\makeatother

\usepackage{bm,cite,algorithm,algorithmicx,float,amsmath,amssymb}
\usepackage[dvips]{graphicx}
\usepackage{subfigure,amsfonts}
\usepackage{mdwmath}
\usepackage{mdwtab}
\usepackage{xcolor}
\usepackage{balance}
\floatname{algorithm}{Algorithm}
\usepackage{titlesec}

\allowdisplaybreaks
\begin{document}


\title{Active-RIS-Aided Covert Communications in NOMA-Inspired ISAC Wireless Systems}

\author{Miaomiao Zhu, Pengxu Chen, Liang Yang, Alexandros-Apostolos A. Boulogeorgos,\\
Theodoros A. Tsiftsis, and Hongwu~Liu
\thanks{M. Zhu, P. Chen, and H. Liu are with the School of Information Science and Electrical Engineering, Shandong Jiaotong University, Jinan 250357, China (e-mail: 1628397290@qq.com, 21208019@stu.sdjtu.edu.cn,  liuhongwu@sdjtu.edu.cn).}
\thanks{L. Yang is with the College of Computer Science and Electronic Engineering, Hunan University, Changsha 410082, China (e-mail: liangy@hnu.edu.cn).}
\thanks{A. A. Boulogeorgos is with the Department of Electrical and Computer Engineering, University of Western Macedonia, 50100 Kozani, Greece (e-mail: aboulogeorgos@uowm.gr).}
\thanks{T. A. Tsiftsis is with the Department of Informatics and Telecommunications, University of Thessaly, 35100 Lamia, Greece (e-mail: tsiftsis@uth.gr).}
}

\maketitle

\setcounter{page}{1}
\begin{abstract}
Non-orthogonal multiple access (NOMA)-inspired integrated sensing and communication (ISAC) facilitates spectrum sharing for radar sensing and NOMA communications, whereas facing privacy and security challenges due to open wireless propagation. In this paper, active reconfigurable intelligent surface (RIS) is employed to aid covert communications in NOMA-inspired ISAC wireless system with the aim of maximizing the covert rate.
Specifically, a dual-function base-station (BS) transmits the superposition signal to sense multiple targets, while achieving covert and reliable communications for a pair of NOMA covert and public users, respectively, in the presence of a warden. To maximize the covert rate, the transmission beamforming at BS and reflection beamforming at active-RIS are jointly designed, subject to quality-of-service (QoS) requirements of NOMA public user, constraint of Cram\'{e}r-Rao bound for multi-target estimations, and covertness level against warden. Two superposition transmission schemes, namely, the transmissions with dedicated sensing signal (w-DSS) and without dedicated sensing signal (w/o-DSS), are respectively considered in the formulations of the joint transmission and reflection beamforming optimization problems. To this end, we develop the efficient methods involving several penalized Dinkelbach transformation to convert the non-concave covert rate maximization problems into feasible ones and propose an alternating optimization algorithm to obtain stationary solutions for the original problems. Numerical results demonstrate that active-RIS-aided NOMA-ISAC system outperforms the passive-RIS-aided and without-RIS counterparts in terms of covert rate and trade-off between covert communication and sensing performance metrics. Finally, the w/o-DSS scheme, which omits the dedicated sensing signal, achieves a higher covert rate than the w-DSS scheme by allocating more transmit power for the covert transmissions, while preserving a comparable multi-target sensing performance.
\end{abstract}


\section{Introduction}

Next generation wireless systems are envisioned with sensing functionalities that will allow them to achieve high-precision positioning of massive devices and real-time reconstructions of the physical world, which are key enablers for anytime, anywhere, internet of everything (IoE), as well as vehicles networking  \cite{ISAC_IoTnetws,ISAC_VCNetw}.
With the advancements of massive multiple-input multiple-output (MIMO) schemes, ultra-wide spectral usage, and ultra-high frequency transmission, such as millimeter and terahertz waves, wireless communication systems naturally possess  
sensing functions with exceptional accuracy and resolution \cite{Dual-functional_Waveform}.
By co-designing communication and sensing systems together in order to use the same hardware and signal processing platforms, same wireless resources, and unified control modules, collaboration efficiencies can be significantly improved \cite{ISAC_IoTnetws}. 
In this direction, novel concepts of integrated sensing and communication (ISAC) have recently sparked heated discussions, particularly on co-designing of dual-function waveforms \cite{Dual-functional_Procoding,CRB_RC}.

As a promising spectral efficiency (SE)-enhancement technique, non-orthogonal multiple access (NOMA) has been applied to enable ubiquitous high-quality communication and high-resolution sensing in NOMA-inspired ISAC wireless systems \cite{NOMA_ISAC,NOMA_RC}, which were frequently constrained by limited wireless resources.
Through superposition coding at the transmitter and successive interference cancellation (SIC) at the receiver, NOMA-ISAC addressed inter-user interference (IUI) effectively, while additional degrees of freedom (DoF) were introduced by adequate dual-functional beamforming \cite{NOMA_ISAC}. To simultaneously deliver multicast and unicast messages, common beamforming and transmit power allocation were jointly optimized in NOMA-ISAC wireless systems with the aim of enhancing radar sensing accuracy and achieving extra DoF for communications \cite{NOMA_RC}. 
Inspired by interference management of NOMA, the dedicated sensing signal (DSS) was partially treated as virtual communication signal and eliminated by SIC at communication receivers \cite{NOMA_inspired_ISAC}. In uplink ISAC systems, a pure-NOMA-based scheme was reported to guarantee the sensing-prior system performance with a fixed communication-to-sensing decoding order \cite{NOMA_ISAC_6G}. As a step further, a semi-NOMA-based scheme was introduced to provide flexible resource allocation strategies between sensing and communication, which effectively satisfied different objectives of sensing and communication \cite{Semi_ISAC_NOMA}.

Due to open and broadcasting natures of wireless propagation, sensitive and privacy information embedded in NOMA-ISAC waveforms was susceptible to interception and eavesdropping \cite{Secure_NOMA_ISAC,Secure_RC_NOMA}. Considering radar target as potential eavesdropper, the authors of \cite{Secure_NOMA_ISAC} exploited artificial jamming and NOMA signal for target detection and eavesdropping prevention. In \cite{Secure_RC_NOMA}, confidential information of NOMA users was concealed in radar beamforming, in which strong interference was conveyed to damage eavesdropping channels. 
By using SIC, artificial jamming and sensing interference were eliminated by NOMA users to enhance communication performance \cite{Secure_NOMA_ISAC,Secure_RC_NOMA}. 
Nevertheless, exposing existences of legitimate transmissions to malicious users brought potential security risks to ISAC systems \cite{6G_ISACS, Covert_ISAC}, e.g., exposing only network traffic pattern can lead to sensitive information leakage. Without additional spectrum investments, it is paramount importance to enhance security performance of ISAC systems by resorting to new technologies. 

Fortunately, reconfigurable intelligent surface (RIS) provided a promising approach to simultaneously facilitate radar sensing and covert communications \cite{RIS_Radar_CovertCom}.
Since RIS possesses ability to manipulate propagation environments intelligently and efficiently, virtual line-of-sight (LoS) links can be created for radar sensing and communication resulting in new DoFs for ISAC optimizations \cite{RIS_ISAC,RIS_DFRC,NOMA_RIS_ISAC}.   
By controlling reflection amplitude and phase-shift of each element of RIS, the reflected wireless signals can be tuned constructively or destructively, such that signal power in desired directions can be strengthened or weakened \cite{smart_reflect-array}. Recently, RIS was utilized to facilitate covert communications in ISAC systems \cite{RIS_Radar_CovertCom} and NOMA systems \cite{Covert_NOMA_IRS,Covert_NOMA_RS_RIS}, which provided a higher level of security than conventional physical layer security (PLS) by concealing legitimate transmissions against wardens.
Compared to conventional covert communications in NOMA systems, where only the signal of a NOMA public user was utilized as a shield for covert transmissions \cite{NOMA_Covert_Channel}, RIS-aided superposition transmissions hide communication behaviors cost-effectively by leveraging RIS's 
phase-shift uncertainties and superposition transmissions of NOMA public and covert users.

Due to ``multiplicative fading'' effects, reflection link qualities of passive-RIS-aided communications were limited and corresponding capacity gains were negligible. To overcome this limitation, active-RIS was introduced in \cite{Active_RIS} and \cite{AvsP_RIS}, which integrated amplifiers into meta-elements to amplify and reflect signals, compensating for substantial path-loss of reflection links and obviating need for complex and power-hungry radio frequency (RF) chains \cite{Active_RIS}. Owning to the aforementioned benefits, PLS  \cite{ARIS_secure} and covert communication performances \cite{Active_IRS_Covert} were improved by optimizing active-RIS reflection beamforming and collaborating with base-station (BS) precoding optimization. In active-RIS-aided multiple-input single-output (MISO) NOMA systems \cite{ARIS_covert_MISO_NOMA},  covert rate was maximized subject to the QoS requirements of NOMA public users and the maximum reflection amplitude and power budget of active-RIS.  

Recently, active-RIS has been applied to enhance communication quality and sensing capability for ISAC systems in which direct links from BS to potential targets have been blocked  \cite{AIRS_MIMO_ISAC,Joint_ARIS_ISAC,Active_IRS_ISAC_C-RAN,HARIS_ISAC}.
To enhance sensing capability for MIMO-ISAC systems, active-RIS was deployed to shape reflection beampattern towards a point-like target \cite{AIRS_MIMO_ISAC}. 
Different from \cite{AIRS_MIMO_ISAC}, in which only communication signal was used to achieve dual-function ISAC, DSS was applied in \cite{Joint_ARIS_ISAC} where transmit beamforming and reflection beamfomring were joint optimized to maximize the radar signal-to-noise ratio (SNR). Furthermore, to enhance the non-line-of-sight (NLoS) sensing capability, active-RIS was deployed to assist remote radio heads in cloud radio access networks (C-RANs) to conduct dual-function ISAC \cite{Active_IRS_ISAC_C-RAN}. 
Taking into account target location estimation errors, a hybrid active-passive-RIS was deployed to maximize the minimum sensing beampattern gain while maintaining a pre-defined signal-to-interference-plus-noise ratio (SINR) for each user \cite{HARIS_ISAC}. The high path-loss was intensively tackled by using active-RIS for ISAC systems, with the aim of not only improving the communication and sensing performances \cite{ARIS_THz_ISAC, Active_RIS_ISAC, Two_stage_ARIS_DFRC}, but also enhancing the PLS capabilities \cite{Active_RIS_ISAC_Secure}. Note that the above active-RIS-aided ISAC contributions, except for  \cite{AIRS_MIMO_ISAC}, have exploited DSS to achieve dual-function ISAC. 

Despite the vast research effort that was put on analyzing and optimizing active-RIS empowered wireless systems, how to deploy active-RIS to make the most out of the covert communication for NOMA-ISAC wireless systems is still unknown. 
Additionally, works in \cite{AIRS_MIMO_ISAC,Joint_ARIS_ISAC,ARIS_THz_ISAC,Active_RIS_ISAC,Active_RIS_ISAC_Secure,Two_stage_ARIS_DFRC} investigated dual-function beamforming with respect to a single target, which was not a typical scenario in practice. Despite multiple targets being considered in \cite{HARIS_ISAC,Active_IRS_ISAC_C-RAN}, the exploited sensing metric, i.e., beampattern gain, was too rough for estimating target parameters. From the perspective of dual-function waveforms, although the usage of DSS and the absence of DSS can improve both the communication and sensing performances, a comprehensive comparison of the impact of the usage of DSS and the absence of DSS on the system performance of active-RIS-aided ISAC systems is missing. In particular, the effects of exploiting DSS in active-RIS-aided NOMA-ISAC systems need to be quantified, especially when DSS is regarded as a part of the NOMA superposition signal. Another question that we need to answer is the following: when active-RIS is deployed to aid covert communications in practical NOMA-ISAC systems, should we apply DSS or not to achieve the enhanced covert communication performance in the presence of multiple targets? 

As an attempt to answer the above open questions, we deploy an active-RIS to aid covert communications in a NOMA-ISAC system, where a dual-function BS serves two NOMA users while sensing multiple moving targets. In more detail, one of the NOMA users is a security-required user (covert user) with a high level of covertness and the other is a QoS-required user (public user) with low or no confidentiality requirements. We examine the use of DSS for multi-target sensing and apply the NOMA public user's signal as the covert medium to hide covert transmissions against warden. Taking into account the realistic assumption of multiple moving targets, we derive Cram\'{e}r-Rao bound (CRB) to characterize the sensing performance for the considered NOMA-ISAC wireless system. The trade-off between covert communication and multi-target sensing is investigated and the impact of the use of DSS is extrapolated. With the aim of maximizing the covert rate, transmission beamforming and reflection beamforming are jointly optimized, while satisfying the CRB constraint on multi-target estimation and ensuring the covertness level against warden. By involving user scheduling and pairing schemes, the considered NOMA-ISAC system is scalable to accommodate multiple NOMA users. 

To the best of our knowledge, this is the first paper investigating covert communications for active-RIS-aided NOMA-ISAC wireless systems in the presence of a warden and multiple moving targets and the first attempt to reveal the impacts of DSS on the covert communication and sensing performances systematically. The key contributions of this paper are summarized as follows:

\begin{itemize}
\item We deploy active-RIS to aid covert communications in a NOMA-ISAC system taking into account two superposition transmission schemes. The first scheme only transmits a NOMA signal without DSS (w/o-DSS scheme), whereas the second scheme transmits  NOMA signal with DSS (w-DSS scheme). To precisely measure the sensing performance with respect to multiple moving targets, we exploit CRB as the target sensing metric for both the w/o-DSS and w-DSS schemes. In the w-DSS scheme, both the sensing and communication signals are utilized to not only estimate the parameters of multiple moving targets, but also hide the covert transmission behaviors from the dual-function BS to the NOMA covert user.
In the w/o-DSS scheme, although only the signal of the NOMA public user is treated as a shield to achieve low-probability-of-detection covert communications for the NOMA covert user, more transmit power can be allocated to enhance covert communication performance.   
\item For both the aforementioned superposition transmission schemes, we formulate the covert rate maximization problems subject to the QoS requirements of the NOMA public user, CRB constraint, power budget constraints at BS and active-RIS, and covertness constraint. We decouple the original non-concave optimization problem into two sub-problems of transmission beamforming optimization at BS and reflection beamforming optimization at active-RIS, respectively, to tackle the optimization variables coupled in the objective and constraints. The resulted rank-one constrained problem  of transmission beamforming optimization is solved with the aid of a penalty-term, while the rank-one constrained fractional programming problem of reflection beamforming optimization is solved by exploiting a penalized Dinkelbach transform. Then, we introduce an alternating optimization (AO) algorithm to optimize the transmission and reflection beamforming in an alternating manner.
\item Extensive numerical results clarify the effectiveness of the derived solutions and AO algorithm. It is demonstrated that the active-RIS-aided NOMA-ISAC system outperforms the passive-RIS-aided and without-RIS counterparts in terms of the covert rate and trade-off between communication and sensing performances because active-RIS amplifies the incident signal to obtain additional covert rate gain and allows BS to allocate more transmit power for target sensing. The impacts of DSS on the ISAC system performance are also revealed by the numerical results. Compared to the w-DSS scheme, the w/o-DSS scheme achieves a higher covert rate because more transmit power is available for covert transmissions by omitting DSS, while exploiting only the NOMA signal ensures the required multi-target sensing performance in the considered NOMA-ISAC system.
\end{itemize}

The remainder of the paper is organized as follows: Section II presents the system model. The covert beamforming
design is specified in Section III. Numerical results are illustrated
in Section IV. Finally, Section V concludes the paper.

{\em Notations:} Throughout the paper, scalars are represented by italic letters, and vectors and matrices are represented by boldfaced lowercase and uppercase letters, respectively. ${\rm{Pr}}\{\cdot\}$ denotes probability, $\mathbb{E}\left\{  \cdot  \right\}$ denotes the expectation, and $\| \cdot \|$ denotes the Euclidean norm of a complex vector.
Superscript $(\cdot)^T$, $(\cdot)^*$ and $(\cdot)^H$ denote vector/matrix transpose, conjugate and Hermitian transpose, respectively. The Hadamard and Kronecker products are denoted by $ \odot$ and $\otimes$, respectively. ${\mathbb{C}^{N \times M}}$ and $ \mathbb{R} ^{N \times M}$ denote the spaces of $N \times M$  complex-valued matrices and real-valued matrices, respectively. $\mathbb{H}^N$ represents the set of all $N\times N$ complex Hermitian matrixes. ${\mathbf{A}} \succeq 0$ indicates that $\mathbf{A}$ is a positive semi-definite matrix. The symbols ${\rm{rank}}({\mathbf{A}})$ and ${\rm tr}({\mathbf{A}})$ denote the rank and trace of matrix $\mathbf{A}$, respectively. $\mathbf{0}_N$ and $\mathbf{1}_N $  denote the $N$-dimensional all-zero and all-ones vectors, respectively. ${{\mathbf{I}}_N}$ denote the $N \times N$ identity matrix.
The notation $\mathcal{C}\mathcal{N}( {\nu ,{\sigma ^2}})$ denotes a circular symmetric complex Gaussian (CSCG) random
distribution with mean $\nu$ and variance $\sigma^2$.

\begin{figure}[tb]
   \begin{center}
    \includegraphics[width=3.3in]{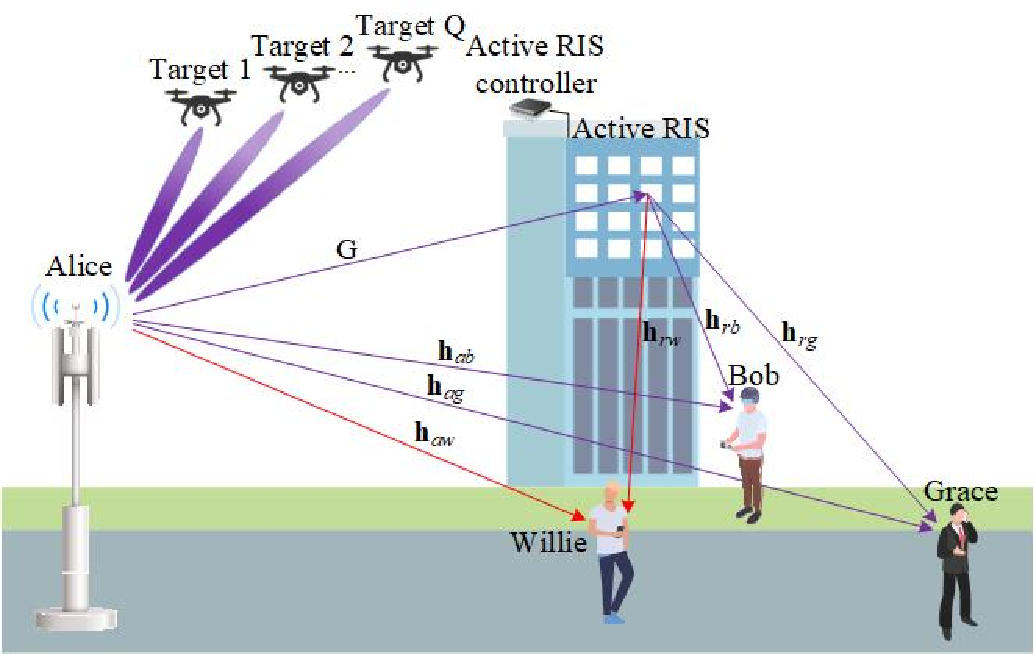}
    \caption{The active-RIS-aided NOMA-ISAC system model.}
   \label{fig:system_model}
   \end{center}
   \vspace{-0.2in}
\end{figure}

\section{System Model}

We consider an active-RIS-aided NOMA-ISAC system consisting of a dual-function BS (Alice), an active-RIS, a warden (Willie), a NOMA public user (Grace), a NOMA covert user (Bob), and multiple moving targets indexed by $\mathcal{Q}=\{1,...,Q\}$, as shown in Fig. \ref{fig:system_model}. We assume that Bob has a high-level of confidentiality and Grace only has reliability requirements, which is a realistic assumption in IoE applications. For example, military departments require covert communications in specific scenarios, while weather forecasts have low or no confidential requirements. 
In the NOMA-ISAC system, Alice is equipped with a uniform linear array (ULA) consisting of $M_\mathrm{t}$ transmit antennas and $M_\mathrm{r}$ receive antennas with half-wavelength spacing. For simplicity, we assume Alice works in full-duplex mode with perfect self-interference cancellation and $M_\mathrm{t}=M_\mathrm{r} =M$. Each one of Grace, Bob, and Willie is equipped with a single antenna. The active-RIS consists of $N$ active reflecting elements and is deployed close to the users to assist information communications.
We assume that targets fly at low altitudes, which results in strong LoS links from Alice to targets and vice versa, while active-RIS is deployed far away from targets resulting in weak RIS reflection links. Thus, we only consider direct links for target sensing \cite{RIS_DFRC,Two_stage_ARIS_DFRC,RIS_ISAC}.
In addition, Willie keeps monitoring the downlink transmission from Alice to the NOMA users and determining whether Alice is transmitting covert information to Bob.
The channel between Alice and node $k$ is denoted by ${{\mathbf{h}}_{ak}} \in {\mathbb{C}^{M \times 1}}$
with $k=g$, $b$, and $w$ standing for Grace, Bob, and Willie, respectively. The Alice-RIS channel and RIS-$k$ channel are denoted by  ${{\mathbf{G}}} \in {\mathbb{C}^{N \times M}}$ and ${{\mathbf{h}}_{{rk}}} \in {\mathbb{C}^{N \times 1}}$, respectively. All of the communication channels are assumed to experience Rician fading with the distance-dependent path-loss, e.g., the Alice-RIS channel $\mathbf{G}$ can be obtained as
\begin{eqnarray}
 \mathbf{G} = \sqrt{\mathcal{L}_{ar}} \Bigg(\sqrt{\frac{\beta_{ar}}{1+\beta_{ar}}} \mathbf{G}^\text{LoS}+  \sqrt{\frac{1}{1+\beta_{ar}}} \mathbf{G}^\text{NLoS}\Bigg),
\end{eqnarray}
where $\mathcal{L}_{ar}$ is the distance-dependent path-loss, $\beta_{ar}$ is the Rician factor, $\mathbf{G}^\text{LoS}$ denotes the LoS component that is expressed as $\mathbf{G}^\text{LoS}=\mathbf{a}_N(\theta_{ra}) \mathbf{a}_M^H(\theta_{ar})$, and $\mathbf{G}^\text{NLoS}$ denotes the Rayleigh fading component with zero mean and unit variance.
Vectors $\mathbf{a}_M(\theta_{ar} )$ and $\mathbf{a}_N(\theta_{ra})$ are the array response vectors of Alice and active-RIS, respectively, which are mathematically given by
$\mathbf{a}_M(\theta_{ar} )= [1, e^{\jmath 2\pi \frac{\delta_1}{\lambda}  \sin(\theta_{ar})}, ..., e^{\jmath 2\pi (M-1) \frac{\delta_1}{\lambda}\sin(\theta_{ar})}]^T$ and $\mathbf{a}_N(\theta_{ra} )= [1, e^{\jmath 2\pi \frac{\delta_2}{\lambda}  \sin(\theta_{ra})}, ..., e^{\jmath 2\pi (N-1) \frac{\delta_2}{\lambda}\sin(\theta_{ra})}]^T$, respectively,
with $\lambda$, $\delta_1 = \lambda/2$, and $\delta_2 = \lambda/2$ being the carrier wavelength, adjacent antenna spacing of Alice, and adjacent element spacing of the active-RIS, respectively.  The Alice-$k$ and RIS-$k$ channels are independent and identical. 
Following the works in \cite{Covert_NOMA_IRS,Covert_NOMA_RS_RIS}, we assume that Alice possesses statistical knowledge of the channel state information (CSI) of the links involving Willie, along with perfect CSI for the other communication channels. Willie, on the other hand, can attain perfect CSI for its monitoring channels. It is assumed that all channels experience quasi-static block fading, remaining constant within each block.
In the rest of this section, two transmission schemes, namely, the w/o-DSS and w-DSS schemes, are respectively presented.

\subsection{w/o-DSS Scheme}
\emph{1) Communication Model:}
In each transmission block, Alice transmits the superimposed NOMA signal to serve two users while sensing the moving targets. Consider $L$ blocks of transmission and radar pulses within one coherent processing interval (CPI) indexed by $\mathcal{L} = [1,\cdots, L]$, the transmitted signal at time index $l$ can be expressed as
\begin{eqnarray}
\mathbf{x}[l] = \mathbf{w}_gs_g[l] +\mathbf{w}_bs_b[l] = \mathbf{W}_c \mathbf{s}_c[l],\label{superimposed_signal}
\end{eqnarray}
where $\mathbf{s}_c[l]=[s_g[l],s_b[l]]^T \in \mathbb{C}^{2\times 1}$ denotes the NOMA communication symbols satisfying $\mathbb{E}\{ \mathbf{s}_c \mathbf{s}_c^H\} = \mathbf{I}_2$, $s_b[l]$ and $s_g[l]$ are the information signals for Bob and Grace, respectively, and $\mathbf{W}_c \triangleq [\mathbf{w}_g,\mathbf{w}_b]\in \mathbb{C}^{M\times 2}$ is the communication beamforming matrix with $\mathbf{w}_p \in \mathbb{C}^{M\times 1} $ denoting the beamforming vector for $s_p[l]$ ($\forall p \in \{g,b\}$). The covariance matrix of the transmitted signal can be calculated by 

\begin{eqnarray}
    \mathbf{R}_x = \frac{1}{L} \sum_{l\in \mathcal{L} }\mathbf{x}[l]\mathbf{x}[l]^H = \mathbf{W}_c\mathbf{W}_c^H.
\end{eqnarray}

In each transmission block, Alice transmits signal to active-RIS. Then, the incident signal is reflected by active-RIS. Taking into account the additive white Gaussian noise (AWGN) at active-RIS, the reflected signal at time index $l$ can be written as
\begin{eqnarray}
\mathbf{y}_r[l] = \mathbf{\Phi } \mathbf{G}\mathbf{x}[l]+\mathbf{\Phi} \mathbf{z}_r[l],
\end{eqnarray}
where $\mathbf{\Phi} \in {\mathbb{C}^{N \times N}}$ is an $N \times N$ diagonal matrix with its $n$th diagonal entity $\mathbf{\Phi}_{n,n}$ denoting the reflection coefficient of the $n$th element of active-RIS and ${\mathbf{z}}_r[l] \sim  \mathcal{C}\mathcal{N}\left( {\mathbf{0}_N, \sigma _r^2 {\mathbf{I}_N}} \right)$ denotes AWGN at the active-RIS.
In particular,  $|\mathbf{\Phi}_{n,n}| \le \eta_n$ and $\arg(\mathbf{\Phi}_{n,n}) \in [0,2\pi )$ respectively represent the reflecting amplitude and phase shift of the $n$th element and $\eta_n > 1 $ is the allowed maximum reflecting amplitude.
In this setup, the total power of the reflected signal is limited by
\begin{eqnarray}
\|\mathbf{\Phi } \mathbf{G} \mathbf{W}_c\|_F^2 +\| \mathbf{\Phi} \|_F^2 \sigma_r^2 \le P_r^\mathrm{max},\label{RIS_power_constraint}
\end{eqnarray}
where $P_r^\mathrm{max}$ is the maximum power budget at the active-RIS.
Then, the received signal at node $k$, $k\in\{g,b,w\}$, at time index $l$, can be expressed as
\begin{eqnarray}
y_{k}[l] = \mathbf{g}_k^H \mathbf{x}[l]+ \mathbf{h}_{{rk}}^H \mathbf{\Phi} \mathbf{z}_r + z_{k}[l],
\end{eqnarray}
where $\mathbf{g}_k^H={{\mathbf{h}}_{ak}^H + {\mathbf{h}}_{{rk}}^H{\mathbf{\Phi}} \mathbf{G}}$ is the equivalent composite channel from Alice to node $k$ and ${z_{k}} \sim \mathcal{C}\mathcal{N}\left( {0,\sigma _{b}^2} \right)$ is the AWGN at node $k$. With the assumption that $\|\mathbf{g}_b\|^2 \ge \|\mathbf{g}_g\|^2$, i.e., Bob and Grace are the near and far users, respectively, Alice allocates more transmit power to transmit $s_g$ rather than $s_b$, i.e., $\|\mathbf{w}_g\|^2 >\|\mathbf{w}_b\|^2$, to guarantee the successfully SIC. In this paper, we assume perfect SIC at Bob and Grace.
According to NOMA principles, Bob detects $s_g$ and removes the corresponding interference by exploiting SIC followed by the detection of its own signal $s_b$. Grace detects $s_g$ by treating the signal related to $s_b$ as noise.
The achievable rates of Bob corresponding to the transmissions of $s_g$ and $s_b$ are given by $ R_{b,{s_g}} = {\log _2} (1 + \gamma_{b, s_g})$ and $ R_{b,{s_b}} = {\log _2} (1 + \gamma_{b, s_b})$, respectively, with the received signal-to-interference-plus noise ratios (SINRs) being given by
\begin{eqnarray}
\gamma_{b, s_g} = \frac{| \mathbf{g}_b^H {\mathbf{w}}_{g} |^2}{
 |\mathbf{g}_b^H{{\mathbf{w}}_{b}} |^2+
 \|\mathbf{h}_{rb}^H \mathbf{\Phi} \| ^2 \sigma _r^2 + \sigma _{b}^2} \label{gamma_b_sg}
\end{eqnarray}
and
\begin{eqnarray}
\gamma_{b, s_b} = \frac{\left| \mathbf{g}_b^H {{\mathbf{w}}_{b}} \right|^2}{ \| \mathbf{h}_{rb}^H \mathbf{\Phi} \| ^2 \sigma _r^2 + \sigma _{b}^2}. \label{gamma_b_sb}
\end{eqnarray}
Moreover, the achievable rate of Grace is given by
\begin{equation}
R_{g,s_g} \!=\! \mathrm{log}_2 \left(\!1\!+\! \frac{| \mathbf{g}_g^H {\mathbf{w}}_{g} |^2}{|\mathbf{g}_g^H{{\mathbf{w}}_{b}} |^2\!+\!
  \| \mathbf{h}_{rg}^H \mathbf{\Phi} \| ^2 \sigma _r^2 \!+\! \sigma _{g}^2}\!\right)\!\!. \label{Rgsg}
\end{equation}
To make SIC feasible at the NOMA receivers, the achievable rate at Grace to detect its own message should be no more than the achievable rate at Bob to detect $s_g$. Thus, we have
\begin{eqnarray}
 R_{b,s_g} \ge  R_{g,s_g}.
\end{eqnarray}

\emph{2) Sensing Model:}
Since the NOMA signals are completely known at Alice, Alice is able to process the echo waves reflected by potential targets for parameters estimation. Therefore, in the considered NOMA-ISAC system, the transmitted NOMA signal can be exploited for target sensing \cite{NOMA_RC}. As typically performed in radar sensing, the Alice-targets links are assumed to be LoS. The echo signal received at Alice at time index $l$ is given by
\begin{eqnarray}
\mathbf{y}_a[l] = \sum_{q \in \mathcal{Q}}  \alpha_q e^{\jmath 2\pi \mathcal{F}_{D_q} l T} \mathbf{h}_{aq} \mathbf{h}^T_{aq} \mathbf{x}[l] +\mathbf{z}_a[l], \label{eacho_signal}
\end{eqnarray}
where $\alpha_q \sim \mathcal{CN}(0, \sigma_q^2)$ is the $q$th moving target's complex reflection factor determined by its radar cross-section (RCS), $\mathcal{F}_{D_q}$  denotes Doppler frequency of the $q$th moving target, $T$ denotes the symbol period, $\mathbf{h}_{aq} = \mathcal{L}_{aq}\mathbf{a}_M(\theta_{aq})$ with $\mathbf{a}_M(\theta_{aq})$ being the transmit steer vector at Alice, and $\mathbf{z}_a[l] \in \mathbb{C}^{M \times 1}$ denotes the AWGN following $\mathcal{CN}(\mathbf{0}_M,\sigma^2 \mathbf{I}_M)$. In particular, $\mathcal{F}_{D_q}=\frac{2v_q f_c}{c}$ with $v_q$ being the velocity of the $q$th moving target, and $c$ and $f_c$ standing for the speed of light and carrier frequency, respectively. Since the considered NOMA-ISAC system conducts target sensing in a monostatic manner,  the direction of arrival (DoA) and the direction of departure (DoD) are the same.
For notation simplicity, \eqref{eacho_signal} is equivalently rewritten as
\begin{eqnarray}
\mathbf{y}_a[l] = \mathbf{A V E}[l] \mathbf{A}^T \mathbf{x}[l] + \mathbf{z}_a[l],
\end{eqnarray}
where  $\mathbf{A}=[\mathbf{h}_{a1}, \ldots, \mathbf{h}_{aQ}]$, $\mathbf{V}=\mathrm{diag}(\bm{\alpha})$ with $\bm{\alpha}= [\alpha_1,\ldots, \alpha_Q]^T$, and $\mathbf{E}[l] \!\!=\!\!\mathrm{diag}([e^{j2\pi \mathcal{F}_{D_1}lT}, \ldots, e^{j2\pi \mathcal{F}_{D_Q}lT}]^T) $.
For the sensing, we are interested in estimating the multiple moving targets's parameters by using CRB, which is the lower bound on the variance of unbiased estimators and equivalent to the inverse of the fisher information matrix (FIM) \cite{CRB_RC}. For target $q$, we define the estimation parameters $\bm{ \xi}_q=[\theta_{aq},\Re(\alpha_q),\Im(\alpha_q),\mathcal{F}_{D_q}]^T$, $\forall q\in \mathcal{Q}$, and the FIM matrix for estimating $\bm{\xi}= \{\bm{\xi}_1,\ldots,\bm{\xi}_Q\}$ of multiple moving targets as $\mathbf{F} \in \mathbb{C}^{4Q \times 4Q}$. As a result, $\mathbf{F}$ can be partitioned as
\begin{equation}
\mathbf{F} =2 \left[ {\begin{array}{*{20}{c}}
  {\Re(\mathbf{F}_{11})}&{\Re(\mathbf{F}_{12})}&{-\Im(\mathbf{F}_{12})}&{-\Im(\mathbf{F}_{14})} \\
  {\Re^T(\mathbf{F}_{12})}&{\Re(\mathbf{F}_{22})}&{-\Im(\mathbf{F}_{22})}&{-\Im(\mathbf{F}_{24})} \\
  {-\Im^T(\mathbf{F}_{12})}&{-\Im^T(\mathbf{F}_{22})}&{\Re(\mathbf{F}_{22})}&{\Re(\mathbf{F}_{24})} \\
  {-\Im^T(\mathbf{F}_{14})}&{-\Im^T(\mathbf{F}_{24})}&{\Re^T(\mathbf{F}_{24})}&{\Re(\mathbf{F}_{44})}
\end{array}} \right] \!, \label{FIM}
\end{equation}
where
\begin{eqnarray}
\mathbf{F}_{11}& \!\!\!=\!\!\!& (\mathbf{\dot{A}}^H\mathbf{Z}^{-1}\mathbf{\dot{A}})\odot
(\mathbf{V}^*\mathbf{A}^H\mathbf{R}^*_x\mathbf{A}\mathbf{V})\odot  \mathbf{\Sigma}_1 \nonumber \\
&\!\!\!\!\!\!&+(\mathbf{\dot{A}}^H\mathbf{Z}^{-1}\mathbf{A})\odot
(\mathbf{V}^*\mathbf{A}^H\mathbf{R}^*_x\mathbf{\dot{A}}\mathbf{V})\odot  \mathbf{\Sigma}_1 \nonumber \\
&\!\!\!\!\!\!&+(\mathbf{A}^H\mathbf{Z}^{-1}\mathbf{\dot{A}})\odot
(\mathbf{V}^*\mathbf{\dot{A}}^H\mathbf{R}^*_x\mathbf{A}\mathbf{V})\odot  \mathbf{\Sigma}_1 \nonumber \\
&\!\!\!\!\!\!&+(\mathbf{A}^H\mathbf{Z}^{-1}\mathbf{A})\odot
(\mathbf{V}^*\mathbf{\dot{A}}^H\mathbf{R}^*_x\mathbf{\dot{A}}\mathbf{V})\odot  \mathbf{\Sigma}_1 , \nonumber \\
\mathbf{F}_{12} & \!\!\!=\!\!\!& (\mathbf{\dot{A}}^H\mathbf{Z}^{-1}\mathbf{A})\odot
(\mathbf{V}^*\mathbf{A}^H\mathbf{R}^*_x\mathbf{A}\mathbf{V})\odot  \mathbf{\Sigma}_1 \nonumber \\
&\!\!\!\!\!\!&+(\mathbf{A}^H\mathbf{Z}^{-1}\mathbf{A})\odot
(\mathbf{V}^*\mathbf{\dot{A}}^H\mathbf{R}^*_x\mathbf{A}\mathbf{V})\odot  \mathbf{\Sigma}_1,\nonumber \\
\mathbf{F}_{14} & \!\!\!=\!\!\!& (\mathbf{\dot{A}}^H\mathbf{Z}^{-1}\mathbf{A})\odot
(\mathbf{V}^*\mathbf{A}^H\mathbf{R}^*_x\mathbf{A}\mathbf{V})\odot  \mathbf{\Sigma}_2 \nonumber \\
&\!\!\!\!\!\!&+(\mathbf{A}^H\mathbf{Z}^{-1}\mathbf{A})\odot
(\mathbf{V}^*\mathbf{\dot{A}}^H\mathbf{R}^*_x\mathbf{A}\mathbf{V})\odot  \mathbf{\Sigma}_2, \nonumber \\
\mathbf{F}_{22} & \!\!\!=\!\!\!& (\mathbf{A}^H\mathbf{Z}^{-1}\mathbf{A})\odot
(\mathbf{A}^H\mathbf{R}^*_x\mathbf{A})\odot  \mathbf{\Sigma}_1,\nonumber \\
\mathbf{F}_{24} & \!\!\!=\!\!\!& (\mathbf{A}^H\mathbf{Z}^{-1}\mathbf{A})\odot
(\mathbf{A}^H\mathbf{R}^*_x\mathbf{A}\mathbf{V})\odot  \mathbf{\Sigma}_2,  \nonumber \\
\mathbf{F}_{44} & \!\!\!=\!\!\!& (\mathbf{A}^H\mathbf{Z}^{-1}\mathbf{A})\odot
(\mathbf{V}^*\mathbf{A}^H\mathbf{R}^*_x\mathbf{A}\mathbf{V})\odot  \mathbf{\Sigma}_3 , \label{F_44}
\end{eqnarray}
with $\mathbf{\dot{A}}=[\frac{\partial\mathbf{h}_{a1}}{\theta_{a1}}, \ldots, \frac{\partial\mathbf{h}_{aQ}}{\theta_{aQ}}]$, $\mathbf{Z}=\sigma_a^2\mathbf{I}_M$, $(\mathbf{\Sigma}_1)_{i,j} =\sum_{l\in \mathcal{L}}e^{\jmath 2\pi (F_{D_j}-F_{D_i})lT}$, $(\mathbf{\Sigma}_2)_{i,j} $ $\sum_{l\in \mathcal{L}}2\pi l Te^{\jmath2\pi (F_{D_j}-F_{D_i})lT}$, and  
$(\mathbf{\Sigma}_3)_{i,j} =$ $\sum_{l\in \mathcal{L}} (2\pi l T)^2e^{\jmath2\pi (F_{D_j}-F_{D_i})lT}$, $\forall i,j \in \mathcal{Q}$. The detailed derivation is provided in Appendix.

\emph{3) Covertness Requirement:}
Based on the received signal, Willie aims to determine whether the information transmission from Alice to Bob is present. The monitoring of Willie involves two hypotheses: ${\cal{H}}_0$ indicates the absence of the information transmission from Alice to Bob  and  ${\cal{H}}_1$ indicates the presence of the information transmission from Alice to Bob. Under ${\cal{H}}_0$ and ${\cal{H}}_1$, the received signals at Willie at time index $l$ can be respectively expressed as
\begin{eqnarray}
        {\mathcal H_0}:~{y_w}[l] = {{\bf{g}}_{{w}}^H} {\bf{w}}_g {s_g}[l]   +{\bf{h}}_{{{r}}w}^H{\bf{\Phi }}{{\bf{z}}_r} + {z_w}[l]
\end{eqnarray}
and
\begin{eqnarray}
     {\mathcal{H}_1}:~{y_w}[l] = {\bf{g}}_{w}^H\mathbf{x}[l]
     + {\bf{h}}_{rw}^H{\bf{\Phi }}{{\bf{z}}_r}
     + {z_w}[l].
\end{eqnarray}
In general, the detection error probability (DEP) is utilized to
measure the performance of Willie’s hypothesis test, which is defined as 
\begin{eqnarray}
    {\zeta }  =   {\Pr} ( {{{\mathcal D}_1} | {{\mathcal H_0}}  } ) + {\Pr} ( {{{\mathcal D}_0} | {{\mathcal H_1}}  }  )
\end{eqnarray}
where ${\Pr} ( {{{\mathcal D}_1} | {{\mathcal H_0}}  } )$ denotes the probability of false alarm, i.e., the rejection of $\mathcal H_0$ when it is true. ${\Pr} ( {{{\mathcal D}_0} | {{\mathcal H_1}}  }  )$ denotes the probability of miss detection, i.e., the acceptance of $\mathcal H_0$ when it is false. ${\mathcal{D}_0}$ and ${\mathcal{D}_1}$ denote Willie's binary decisions endorsing ${\mathcal{H}_0}$ and ${\mathcal{H}_1}$, respectively. To  
evaluate the DEP, we assume the apriori probabilities are equal in the statistical hypothesis test.
Assuming Willie uses the Neyman-Pearson criterion to detect the covert transmission, the optimal decision rule for minimizing the DEP is the likelihood ratio test, i.e.,
\begin{eqnarray}
\frac{{{p_1}({y_{{w}}})}}{{{p_0}({y_{{w}}})}}\mathop  \gtrless \limits_{{\mathcal{D}_0}}^{{\mathcal{D}_1}} 1, \label{eq:likelihood_ratio}
\end{eqnarray}
where 
\begin{eqnarray}
    {p_0}({y_w}) = \frac{1}{\pi \sigma _0^2} e^{ - {|y_w|^2}/{ \sigma _0^2 }}
\end{eqnarray}
and 
\begin{eqnarray}
    {p_1}({y_w}) = \frac{1}{\pi \sigma _1^2} e^{ -  {|y_w|^2}/{{\sigma _1^2}}} 
\end{eqnarray}
denote the likelihood functions for Willie's received signals under ${\cal{H}}_0$ and ${\cal{H}}_1$, respectively, with $\sigma _0^2 =   | {{\bf{g}}_{{w}}^{H}{\bf{w}}_g}  |^2 + {\left\| {{\bf{h}}_{rw}^H{\bf{\Phi }}} \right\|^2}\sigma _r^2 + \sigma _w^2$ and
$\sigma _1^2 =   \| {{\bf{g}}_{{w}}^{H}{\bf{W}}_c} \|^2 + { \| {{\bf{h}}_{rw}^H{\bf{\Phi }}}  \|^2}\sigma _r^2 + \sigma _w^2$, respectively. With respect to Willie's received signal power, the optimal decision rule for Willie can be rewritten as
\begin{eqnarray}
P_w \mathop  \gtrless \limits_{{\mathcal{D}_0}}^{{\mathcal{D}_1}} {\tau ^*},\label{eq:optimal_tau}
\end{eqnarray}
where
$ P_w \mathop = \limits^{ L \to \infty} \frac{1}{L} \sum\nolimits_{l \in \mathcal{L}} |y_w[l]|^2 $ and ${\tau ^ * } = \frac{{\sigma _0^2\sigma _1^2}}{{\sigma _0^2 - \sigma _1^2}}\ln \frac{{\sigma _1^2}}{{\sigma _0^2}} > 0$ denotes the optimal detection threshold.
Under the optimal detection rule, we can derive the minimum DEP achieved by Willie corresponding to the two hypotheses, as in \cite{Covert_AWGN_channel}:
\begin{eqnarray}
{\zeta ^*} &\!\!\! = \!\!\!&  {\Pr} ( {{{\mathcal D}_1} | {{\mathcal H_0}}  } ) + {\Pr} ( {{{\mathcal D}_0} | {{\mathcal H_1}}  }  )  \nonumber  \\
&\!\!\! = \!\!\!& {\Pr} ( { P_w \ge \tau^* | {{\mathcal H_0}}  } ) + {\Pr} ( { P_w < \tau^* | {{\mathcal H_1}}  }  )  \nonumber  \\
    &\!\!\! = \!\!\!&  1 + {\left( {\frac{{\sigma _1^2}}{{\sigma _0^2}}} \right)^{ - \frac{{\sigma _1^2}}{{\sigma _1^2 - \sigma _0^2}}}} - {\left( {\frac{{\sigma _1^2}}{{\sigma _0^2}}} \right)^{ - \frac{{\sigma _0^2}}{{\sigma _1^2 - \sigma _0^2}}}} . \label{eq:detection_error_probability}
\end{eqnarray}
To facilitate the design of covert communications, we further introduce a lower bound on $\zeta^*$ \cite{Covert_AWGN_channel}
\begin{eqnarray}
{\zeta ^*} &\!\!\! \ge \!\!\!& 1 - \sqrt {\frac{1}{2}{\cal D}({p_0}({y_w})||{p_1}({y_w}))},~~~~ \label{eq:lower_bound}
\end{eqnarray}
where ${\cal D}({p_0}({y_w})||$ ${p_1}({y_w})) $ refers to the Kullback-Leibler (KL) divergence, which can be calculated as:
\begin{eqnarray}
{\cal D}({p_0}({y_w})||{p_1}({y_w})) &\!\!\! = \!\!\!& \ln \left(\frac{\sigma_1^2}{\sigma_0^2}\right) +  \frac{\sigma_0^2}{\sigma_1^2}  -1 . \label{eq:KL_divergence}
\end{eqnarray}
To ensure covertness, the minimum DEP of Willie should satisfy ${\zeta^*} \ge 1 - \varepsilon$, where $\varepsilon > 0$ is the desired level of covertness.
In this work, we set a tighter constraint, i.e., 
\begin{eqnarray}
{\mathcal D}({p_0}({y_w}) | {{p_1}({y_w})} ) \le 2{\varepsilon ^2},
\end{eqnarray}
to ensure covertness, taking into account the lower bound on $\zeta^*$ in \eqref{eq:lower_bound}.
According to \eqref{eq:KL_divergence}, by applying the monotonicity of the function $f(x) = \ln x  + \frac{1}{x} - 1$ in the interval $[1, \infty)$, the covertness constraint can be reformulated as:
\begin{eqnarray}
 |\mathbf{g}_w^H\mathbf{w}_b|^2 \!+\!(1 \!-\! \kappa)
\Big( |\mathbf{g}_w^H{{\mathbf{w}}_{g}} |^2\!+\!
\|{\bf{h}}_{rw}^H{\bf{\Phi }}\|^2\sigma _r^2\Big)\!\le\!(\kappa \!-\!1)\sigma _w^2,  \label{eq:covertness}
\end{eqnarray}
where $\kappa$ is the unique root of $f(\lambda ) = 2{\varepsilon ^2}$ in the interval $[1, \infty)$. The above covertness constraint indicates that the AWGN introduced by the active-RIS and NOMA public user's signal can be utilized as the shield of covert transmission to enhance the covert performance of the active-RIS aided NOMA-ISAC system.

\subsection{w-DSS Scheme}

In this section, we apply the w-DSS scheme in the considered active-RIS-aided NOMA-ISAC system, in which the BS has the DSS.

\emph{1) Communication Model:} Alice transmits the superimposed  communication and DSSs for information transmission and targets estimation, in which the transmission of the DSS is assumed to be a general multi-beam transmission. Then, the transmitted signal at time index $l$ from Alice is expressed as
\begin{eqnarray}
\mathbf{\bar x}[l] =\mathbf{W}_c \mathbf{s}_c[l]+\mathbf{W}_s \mathbf{s}_s[l]=\mathbf{W}\mathbf{s}[l],\label{superimposed_signal1}
\end{eqnarray}
where ${\mathbf{s}}_s \in \mathbb{C}^{M\times 1}$ is the DSS satisfying $\mathbf{s}_s[l] \mathbf{s}_s^H[l] = \mathbf{I}_M$ and $\mathbf{W}_s \in \mathbb{C}^{M \times M}$ is the sensing beamforming matrix. In \eqref{superimposed_signal1}, $\mathbf{W} \triangleq [\mathbf{W}_c,\mathbf{W}_s]\in \mathbb{C}^{M\times(2+M)}$ is the overall beamforming matrix to be designed and $\mathbf{s}[l]=[\mathbf{s}_c^T[l],\mathbf{s}_s^T[l]]^T\in \mathbb{C}^{(2+M) \times 1}$. We assume that all the signals are independent and pairwise uncorrelated. Therefore, the covariance matrix of the transmitted signal can be calculated as
\begin{eqnarray}
    \mathbf{\bar R}_x = \frac{1}{L} \sum_{l\in \mathcal{L} }\mathbf{\bar x}[l]\mathbf{\bar x}[l]^H = \mathbf{W}\mathbf{W}^H.
\end{eqnarray}
The reflected signal by the active-RIS at time index $l$ can be written as
\begin{eqnarray}
\mathbf{\bar y}_r[l] = \mathbf{\Phi } \mathbf{G}\mathbf{\bar x}[l]+\mathbf{\Phi} \mathbf{z}_r[l].
\end{eqnarray}
Accordingly, the total power of the reflected signal at the active-RIS is limited by
\begin{eqnarray}
\|\mathbf{\Phi } \mathbf{G} \mathbf{W}\|_F^2 +\| \mathbf{\Phi} \|_F^2 \sigma_r^2 \le P_r^\mathrm{max},\label{RIS_power_constraint1}
\end{eqnarray}
Then, the received signal at node $k$, $k\in\{g,b,w\}$ at time index $l$ is given by
\begin{eqnarray}
{\bar y}_{k}[l] = \mathbf{g}_k^H \mathbf{\bar x}[l]+ \mathbf{h}_{{rk}}^H \mathbf{\Phi} \mathbf{z}_r + z_{k}[l]. \label{y_k1}
\end{eqnarray}
It is observed in \eqref{superimposed_signal1} and \eqref{y_k1} each NOMA user suffers  interference induced by information signal of another NOMA user and   DSS. We assume that $\mathbf{s}_s$ is predetermined sequences and the communication users can know it prior \cite{OptiBeam_ISAC}.
Thus, we consider the receivers of the NOMA users have the capability of cancelling interference resulted by DSS before detecting its desirable information signal.
Consequently, Bob detects $s_g$, with the achievable rate as
\begin{eqnarray}
\bar{R}_{b,s_g} = \mathrm{log}_2 \left(1+ \frac{| \mathbf{g}_b^H {\mathbf{w}}_{g} |^2}{|\mathbf{g}_b^H{{\mathbf{w}}_{b}} |^2
  + \| \mathbf{h}_{rb}^H \mathbf{\Phi} \| ^2 \sigma _r^2 + \sigma _{b}^2}\right). \label{Rbsg1}
\end{eqnarray}
After detecting and removing signal of Grace, Bob detects its own signal $s_b$ with the achievable rate that is given by 
\begin{eqnarray}
\bar{R}_{b,s_b} = \mathrm{log}_2 \left(1+ \frac{| \mathbf{g}_b^H {\mathbf{w}}_{b} |^2}{ \| \mathbf{h}_{rb}^H \mathbf{\Phi} \| ^2 \sigma _r^2 + \sigma _{b}^2}\right). \label{Rbsb1}
\end{eqnarray}
Grace decodes its own message by regarding the signal of Bob as interference. Thus, the achievable rate of Grace can be written as
\begin{eqnarray}
\bar{R}_{g,s_g} = \mathrm{log}_2 \left(1+ \frac{| \mathbf{g}_g^H {\mathbf{w}}_{g} |^2}{|\mathbf{g}_g^H{{\mathbf{w}}_{b}} |^2
  + \| \mathbf{h}_{rg}^H \mathbf{\Phi} \| ^2 \sigma _r^2 + \sigma _{g}^2}\right). \label{Rgsg1}
\end{eqnarray}
Similarly, we have $\bar R_{b,s_g} \ge \bar R_{g,s_g}$ to make SIC feasible at the NOMA receiver.

\emph{2) Sensing Model:}
Based on the signal model that both the NOMA communication signal and the DSS are exploited for targets parameters estimation, the echo signal received at Alice at time index $l$ is defined as
\begin{eqnarray}
\mathbf{\bar y}_a[l] = \mathbf{A V E}[l] \mathbf{A}^T \mathbf{\bar x}[l] + \mathbf{z}_a[l],\label{eacho_signal1}
\end{eqnarray}
To estimate $\bm{\xi}$ of multiple moving targets, the expression for the FIM $\mathbf{\bar F}$ 
can be presented in the same structure as that in \eqref{FIM} containing the following entries:
\begin{eqnarray}
\mathbf{\bar F}_{11}& \!\!\!=\!\!\!& (\mathbf{\dot{A}}^H\mathbf{Z}^{-1}\mathbf{\dot{A}})\odot
(\mathbf{V}^*\mathbf{A}^H\mathbf{\bar R}^*_x\mathbf{A}\mathbf{V})\odot  \mathbf{\Sigma}_1 \nonumber \\
&\!\!\!\!\!\!&+(\mathbf{\dot{A}}^H\mathbf{Z}^{-1}\mathbf{A})\odot
(\mathbf{V}^*\mathbf{A}^H\mathbf{\bar R}^*_x\mathbf{\dot{A}}\mathbf{V})\odot  \mathbf{\Sigma}_1 \nonumber \\
&\!\!\!\!\!\!&+(\mathbf{A}^H\mathbf{Z}^{-1}\mathbf{\dot{A}})\odot
(\mathbf{V}^*\mathbf{\dot{A}}^H\mathbf{\bar R}^*_x\mathbf{A}\mathbf{V})\odot  \mathbf{\Sigma}_1 \nonumber \\
&\!\!\!\!\!\!&+(\mathbf{A}^H\mathbf{Z}^{-1}\mathbf{A})\odot
(\mathbf{V}^*\mathbf{\dot{A}}^H\mathbf{\bar R}^*_x\mathbf{\dot{A}}\mathbf{V})\odot  \mathbf{\Sigma}_1 , \nonumber \\
\mathbf{\bar F}_{12} & \!\!\!=\!\!\!& (\mathbf{\dot{A}}^H\mathbf{Z}^{-1}\mathbf{A})\odot
(\mathbf{V}^*\mathbf{A}^H\mathbf{\bar R}^*_x\mathbf{A}\mathbf{V})\odot  \mathbf{\Sigma}_1 \nonumber \\
&\!\!\!\!\!\!&+(\mathbf{A}^H\mathbf{Z}^{-1}\mathbf{A})\odot
(\mathbf{V}^*\mathbf{\dot{A}}^H\mathbf{\bar R}^*_x\mathbf{A}\mathbf{V})\odot  \mathbf{\Sigma}_1,\nonumber \\
\mathbf{\bar F}_{14} & \!\!\!=\!\!\!& (\mathbf{\dot{A}}^H\mathbf{Z}^{-1}\mathbf{A})\odot
(\mathbf{V}^*\mathbf{A}^H\mathbf{\bar R}^*_x\mathbf{A}\mathbf{V})\odot  \mathbf{\Sigma}_2 \nonumber \\
&\!\!\!\!\!\!&+(\mathbf{A}^H\mathbf{Z}^{-1}\mathbf{A})\odot
(\mathbf{V}^*\mathbf{\dot{A}}^H\mathbf{\bar R}^*_x\mathbf{A}\mathbf{V})\odot  \mathbf{\Sigma}_2, \nonumber \\
\mathbf{\bar F}_{22} & \!\!\!=\!\!\!& (\mathbf{A}^H\mathbf{Z}^{-1}\mathbf{A})\odot
(\mathbf{A}^H\mathbf{\bar R}^*_x\mathbf{A})\odot  \mathbf{\Sigma}_1,\nonumber \\
\mathbf{\bar F}_{24} & \!\!\!=\!\!\!& (\mathbf{A}^H\mathbf{Z}^{-1}\mathbf{A})\odot
(\mathbf{A}^H\mathbf{\bar R}^*_x\mathbf{A}\mathbf{V})\odot  \mathbf{\Sigma}_2  \nonumber \\
\mathbf{\bar F}_{44} & \!\!\!=\!\!\!& (\mathbf{A}^H\mathbf{Z}^{-1}\mathbf{A})\odot
(\mathbf{V}^*\mathbf{A}^H\mathbf{\bar R}^*_x\mathbf{A}\mathbf{V})\odot  \mathbf{\Sigma}_3.
\end{eqnarray}

\emph{3) Covertness Requirement:}
In the w-DSS scheme, the received signals at Willie under ${\cal{H}}_0$ and ${\cal{H}}_1$ can be respectively expressed as
\begin{eqnarray}
        {\mathcal H_0}:~{\bar y_w}[l] = {{\bf{g}}_{{w}}^H} \left({\bf{w}}_g {s_g}[l] \!+\! \mathbf{W}_s\mathbf{s}_s[l] \right) \!+\!{\bf{h}}_{{{r}}w}^H{\bf{\Phi }}{{\bf{z}}_r} \!+\! {z_w}[l]
\end{eqnarray}
and
\begin{eqnarray}
     {\mathcal{H}_1}:~{\bar y_w}[l] = {\bf{g}}_{w}^H\mathbf{\bar x}[l]
     + {\bf{h}}_{rw}^H{\bf{\Phi }}{{\bf{z}}_r}
     + {z_w}[l].
\end{eqnarray}
Similar to the derivation of \eqref{eq:covertness}, the covertness constraint under this setup can be formulated as
\begin{eqnarray}
 |\mathbf{g}_w^H\mathbf{w}_b|^2 \!+\!(1 \!-\! \kappa)&\!\!\!\!\!\!
\Big( |\mathbf{g}_w^H{{\mathbf{w}}_{g}} |^2\!+\!
 \| \mathbf{g}_w^H{{\mathbf{W}}_{s}}\|^2 \!+\! \|{\bf{h}}_{rw}^H{\bf{\Phi }}\|^2\sigma _r^2\Big)\!\!\!\!\!\!& \nonumber \\
&\!\!\!\le(\kappa -1)\sigma _w^2.\!\!\!&  \label{eq:covertness1}
\end{eqnarray}
In addition to the AWGN and NOMA public user's signal, the expression in \eqref{eq:covertness1}  indicates that the DSS is also utilized as the covert medium to facilitate covert communication of the NOMA-ISAC system.

\section{Problem Formulation and Proposed Solutions}

In this section, we focus on the joint optimization of the transmission beamforming and reflection beamforming in the active-RIS-aided NOMA-ISAC system.
We firstly formulate the optimization problems when Alice performs the target sensing and communication simultaneously in the two transmission schemes. Then, we present solutions to the formulated optimization problems by designing an AO algorithm for optimizing the transmission beamforming at Alice and reflection beamforming at the active-RIS.

\subsection{Problem formulation}
 
The diagonal elements of the CRB matrix represent the minimum variance of the multiple targets parameters estimated by the unbiased estimator \cite{Estimation_Theory}. 
Our objective is to maximize the covert rate at Bob while limiting CRB. To investigate the tradeoff of covert transmission and target estimation, we jointly optimize the transmission beamforming and the reflection beamforming subject to the power budget constraints of Alice and active-RIS, achievable rate constraint for NOMA, Grace's QoS requirement, CRB constraint and pre-defined covertness level.
Considering the additional constraints imposed by the active-RIS-aided NOMA-ISAC system, the optimization problem without the dedicated sensing sihgnal is formulated as follows:
\begin{subequations}
\begin{align}
({\text{P}}1):~&\mathop {\max }\limits_{ \mathbf{W}_c, \mathbf{\Phi}}~
{ R_{b,s_b} }&
\label{objective1}  \\
 {\text{s.t.}}~~&\|\mathbf{W}_c\|_F^2 \le P_a^{\max },&
\label{P1_1b}\\
& { R_{b,{s_g}}} \ge {\bar R_{g,{s_g}}},~\|\mathbf{w}_g\|^2 \ge \|\mathbf{w}_b\|^2,&
\label{P1_1c}\\
& { R_{g,{s_g}}} \ge R_g^{\min },&
 \label{P1_1d}\\
 & \mathrm{tr}(\mathbf{ F}^{-1} ) \le \mu,&
 \label{P1_1e}\\
&   | {\bf \Phi}_{{n,n}}  | \le {\eta _{n}}, \forall n, & \label{P1_1f}\\
&\eqref{RIS_power_constraint}, ~\eqref{eq:covertness},&\label{P1_1g}
 \end{align}
\end{subequations}
where $\mu$ is the CRB threshold, $P_a^{\max}$ is the maximum transmit power budget at Alice, and $R_g^{\min}$ is the minimum target rate of Grace.
In problem (P1),  \eqref{P1_1b} is Alice's maximum transmit power constraint, \eqref{P1_1c} guarantees successful SIC at Bob and Grace, \eqref{P1_1d} ensures that the QoS requirements of Grace is satisfied, \eqref{P1_1e} represents the CRB constraint, \eqref{P1_1f} indicates the limitation of the maximum reflecting amplitude of $n$th element of ${\bf \Phi}$, and constraints in \eqref{P1_1g} denote reflection power budget at the active-RIS and covertness level, respectively.

For the w-DSS scheme, the corresponding optimization problem is formulated as follows:
\begin{subequations}
\begin{align}
({\text{P}}2):~&\mathop {\max }\limits_{ {\bf W, \Phi}}~
{\bar R_{b,s_b}} &
\label{objective2}  \\
 {\text{s.t.}}~~&\|\mathbf{W}\|_F^2 \le P_a^{\max },&
\label{P2_1b}\\
&{\bar R_{b,{s_g}}} \ge {\bar R_{g,{s_g}}}, ~\|\mathbf{w}_g\|^2 \ge \|\mathbf{w}_b\|^2,&
\label{P2_1c}\\
& {\bar R_{g,{s_g}}} \ge R_g^{\min },&
 \label{P2_1d}\\
&  \mathrm{tr}(\mathbf{\bar F}^{-1})  \le \mu,&
 \label{P2_1e}\\
&   | {\bf \Phi}_{{n,n}}  | \le {\eta _{n}}, ~\forall n,& \label{P2_1f}\\
&  \eqref{RIS_power_constraint1}, ~\eqref{eq:covertness1}, ~\eqref{P1_1f}. \label{P2_1g}
 \end{align}
\end{subequations}
In problems (P1) and (P2), the optimization objectives \eqref{objective1} and \eqref{objective2} are non-concave and the majority of the constraints are non-convex, both problems (P1) and (P2) should be transformed into a concave one to be solved.
 
\subsection{Proposed Solution to Problem (P1)}

The main challenge for solving problem (P1) is the coupling between the optimization variables ${\bf W }_c$ and $\bf \Phi$, which makes it infeasible to jointly optimize ${\bf W}_c$ and $\bf \Phi$. To address this challenge, we decouple problem (P1) into two sub-problems of optimizing the transmission beamforming $\mathbf{W}_c$ and reflection beamforming $\mathbf{\Phi}$, respectively. Furthermore, we design an AO algorithm to obtain the optimal solution of problem (P1). The details of our proposed solution are presented below.

\emph{1) Transmission Beamforming Optimization:} For any given feasible $\mathbf{\Phi}$, problem (P1) can be reduced to optimize ${\bf W}_c$ only. To tackle the non-convex CRB constraint, we transform it into a trackable form by introducing an auxiliary variable $\mathbf{J} \in \mathbb{C}^{4Q \times 4Q}$, $ \mathbf{J} \succeq 0$. Considering that the FIM $\mathbf{F}$ is a positive semidefinite matrix and the function $\mathrm{tr}(\mathbf{A}^{-1})$ is matrix decreasing on the positive semidefinte matrix space \cite[Example 3.46]{Convex_Optimization}, the CRB constraint \eqref{P1_1e} is equivalent to the following two constraints:
\begin{subequations}
\begin{align}
\mathrm{tr}(\mathbf{J}^{-1})  \le  \mu,  \\
\mathbf{F}  \succeq  \mathbf{J}. ~~~
\end{align}
\end{subequations}
To facilitate solving the transmission beamforming optimization problem, we introduce the auxiliary variables $\mathbf{W}_p= \mathbf{w}_p\mathbf{w}_p^H$, $p\in\{g,b\}$, $\mathbf{\Upsilon}_k=\mathbf{g}_k\mathbf{g}_k^H$,  $\mathbf{H}_{rk}=\mathbf{h}_{rk}\mathbf{h}_{rk}^H$, $\mathbf{\Psi}=\mathbf{\Phi}\mathbf{\Phi}^H$, and $\mathbf{\Gamma} = \bar{\mathbf{\Gamma}}^H\bar{\mathbf{\Gamma}}$ with $\bar{\mathbf{\Gamma}}=\mathbf{\Phi}\mathbf{G}$.
One can obtain that ${\left| {\mathbf{g}_k^H {{\mathbf{w}_p}} } \right|^2}= \mathrm{tr}(\mathbf{\Upsilon}_k\mathbf{W}_p)$,  $\|\mathbf{h}_{rk}^H\mathbf{\Phi}\|^2=\mathrm{tr}(\mathbf{H}_{rk}
\mathbf{\Psi})$, and $\|\mathbf{\Phi } \mathbf{G} \mathbf{W}_c\|_F^2 + \| \mathbf{\Phi} \|_F^2 \sigma_r^2 = \mathrm{tr}(\mathbf{\Gamma} \tilde{\mathbf{ W}}_c)+\mathrm{tr}(\mathbf{\Psi})\sigma_r^2$ with $\tilde{\mathbf{ W}}_c = \mathbf{W}_g+\mathbf{W}_b$.
In addition, constraint ${R_{b,{s_g}}} \ge {R_{g,{s_g}}}$ is equivalent to $\|\mathbf{g}_b\|^2 \ge \|\mathbf{g}_g\|^2$, which is independent of the transmission beamforming $\mathbf{W}_c$. Since the $\mathrm{log}(\cdot)$ function is monotonically increasing, maximizing $R_{b,s_b}$ is equivalent to maximizing $|\mathbf{g}_b^H\mathbf{w}_b|^2$. Therefore, the optimization problem with respect to ${\bf W}$ can be formulated as:
\begin{subequations}
\begin{align}
~~~\!\!\!\!\!\!\!\!({\text{P}}3):& \mathop {\max }\limits_{ \mathbf{W}_g, \mathbf{W}_b, \mathbf{J}}
\mathrm{tr}(\mathbf{\Upsilon}_b\mathbf{W}_b)&
\label{P3_a}\\
 {\text{s.t.}}~~&\mathrm{tr}(\tilde{\mathbf{{W}}}_c) \le P_a^{\max },&\label{P3_b}\\
 &\mathrm{tr}(\mathbf{W}_g )\ge \mathrm{tr}(\mathbf{W}_b ),&
\label{P3_c}\\
\!\!\!\!\!&\mathrm{tr}(\mathbf{\Upsilon}_g \mathbf{W}_g) \ge \gamma_\mathrm{th} \Big( \mathrm{tr}(\mathbf{\Upsilon}_g \mathbf{W}_b)+\mathrm{tr}(\mathbf{H}_{rg}
\mathbf{\Psi}) \sigma _r^2 \!+\! \sigma _{g}^2 \Big) ,&
 \label{P3_d}\\
 \!\!\!\!\!& \mathrm{tr}(\mathbf{J}^{-1}) \le \mu,~ \mathbf{F}(\tilde{\mathbf{W}}_c)-\mathbf{J} \succeq 0,&
 \label{P3_e}\\
&\mathrm{tr}(\mathbf{\Gamma} \tilde{\mathbf{{W}}}_c)+ \mathrm{tr}(\mathbf{\Psi})\sigma_r^2 \le P_r^\mathrm{max},&
\label{P3_f}\\
&\mathrm{tr}(\mathbf{\Upsilon}_w\mathbf{W}_b)\!+\!(1\!-\!\kappa)
\Big(\mathrm{tr}(\mathbf{\Upsilon}_w \mathbf{W}_g\! )+\mathrm{tr}(\mathbf{H}_{rw}\mathbf{\Psi})\sigma_r^2\Big)& \nonumber\\
 &~~~~~~~~~~~~~~\qquad\qquad\quad \le (\kappa\!-\!1)\sigma_w^2 ,&\label{P3_g}\\
 & \mathbf{W}_g \succeq 0, ~\mathbf{W}_b \succeq 0, ~\mathbf{J} \succeq 0, & \label{P3_h}\\
& \mathrm{rank}(\mathbf{W}_g) =1,~\mathrm{rank}(\mathbf{W}_b) =1.&\label{P3_i}
\end{align}
\end{subequations}
In problem (P3), constraint \eqref{P3_c} represents the QoS requirements of Grace with $\gamma_\mathrm{th}=2^{R^{\mathrm{min}}_g} - 1$.
To address the rank-one constraint \eqref{P3_i}, we introduce a penalty term $\frac{1}{\iota_1} \sum_{p \in \{ g,b\} }\big( \|\mathbf{W}_p\|_*+\widehat{\mathbf{W}}^{(t)}_p\big)$ into \eqref{P3_a}, where $\iota_1$ is a penalty factor, $\| \cdot \|_*$ denotes the nuclear norm, and $\widehat{{\mathbf{W}}}^{(t)}_p$ is the convex upper bound of the non-convex term $-\|\mathbf{W}_p\|_2$ with  $\|\cdot\|_2$ standing for the spectral norm, i.e., $-\|{\mathbf{W}}_p\|_2 \le \widehat{\mathbf{W}}_p^{(t)}$. One can obtain the convex upper bound $\widehat{\mathbf{W}}_p^{(t)}$ by leveraging the first-order Taylor expansion at the point $\mathbf{W}_p^{(t)}$, which is expressed as
$\widehat{\mathbf{W}}_p^{(t)}\triangleq -\|\mathbf{W}_p^{(t)}\|_2 -\mathrm{tr}[ \mathbf{q}_{\mathrm{max},p}^{(t)} (\mathbf{q}_{\mathrm{max}, p}^{(t)})^H(\mathbf{W}_p -\mathbf{W}_p^{(t)})]$,
where $\mathbf{W}_i^{(t)}$ is the solution obtained in the $t$th iteration and $\mathbf{q}_{\mathrm{max},p}^{(t)} $ is the eigenvector corresponding to the largest eigenvalue of $\mathbf{W}_p^{(t)}$. With the added penalty term, the transmission beamforming optimization problem can be formulated as
\begin{subequations}
\begin{align}
({\text{P}}3.1):& \mathop {\max }\limits_{ \mathbf{W}_g, \mathbf{W}_b,  \tilde{\mathbf{{W}}}_s, \mathbf{J}}
\mathrm{tr}(\mathbf{\Upsilon}_b\mathbf{W}_b)&  \nonumber \\
&~~~~~ \qquad \qquad \!-\!\frac{1}{\iota_1}\!\! \sum\limits_{p \in \{ g,b\} }\!\! \Big(\|\mathbf{W}_p\|_*\!+\!\widehat{\mathbf{W}}_p^{(k)}\Big)&
\label{P3.1_a}\\
 &~~~~\quad{\text{s.t.}}~\eqref{P3_b}-\eqref{P3_h}.&
\label{P3.1_b}
\end{align}
\end{subequations}
Here, if $\iota_1 \to 0$, the exactly rank-one metrics can be guaranteed by maximizing the new objective due to the fact that $\mathrm{rank}(\mathbf{W}_p) = 1$ is equivalent to $\|\mathbf{W}_p\|_*- \|\mathbf{W}_p\|_2 =0$.
It is evident that problem (P3.1) is a concave optimization problem and can be readily solved using existing convex optimization solvers such as CVX.

\emph{2) Reflection Beamforming Optimization:}
For any given ${\bf W}_c$, problem (P1) can be reduced to optimize $\mathbf{\Phi}$ only. To handle the non-concavity of the objective function and constraints \eqref{P1_1c}, \eqref{P1_1d}, and \eqref{P1_1f} with respect to $\bf \Phi$ and formulate a tractable covert rate maximization problem, we first introduce $\mathbf{\bar{u}}=[\phi_1,\ldots,\phi_N]$ with $\phi_n$, $\forall n \in \{1,\ldots,N\}$, being the $n$th diagonal element in $\mathbf{\Phi}$
and $\mathbf{\bar{\Lambda}}_{k,p}=[\mathrm{diag}({{\mathbf{h}}_{{rk}}^H})
{\mathbf{G}} \mathbf{w}_p;\mathbf{h}_{ak}^H \mathbf{w}_p]$ and construct
$\mathbf{\Lambda}_{k,p}= \mathbf{\bar{\Lambda} }_{k,p} \mathbf{\bar{\Lambda}}_{k,p}^H$
and ${\bf U} = {\bf{ u}}{{{\bf{ u}}}^H}$ with ${\mathbf{u}}= [\mathbf{\bar{u}},1]^H$, which results in the expression ${{{\left| {\mathbf{g}_k^H{{\mathbf{w}}_{p}}} \right|}^2}}=\mathrm{tr}(\mathbf{\Lambda}_{k,p}\mathbf{U})$.
Then, by applying the change of variables
$\mathbf{\Omega}_k= \mathrm{diag}\left( |[\mathbf{h}_{rk}]_1|^2,...,
|[\mathbf{h}_{rk}]_N|^2, 0 \right)$,
$\mathbf{\Pi} = \mathrm{diag}([\mathbf{1}_{ N},0]^T)$, and $\mathbf{S}_p= \mathrm{diag}\left( |[\mathbf{s}_p]_1|^2,...,
|[\mathbf{s}_p]_N|^2, 0 \right)$ with $\mathbf{s}_p=
\mathbf{G}\mathbf{w}_p$,
we obtain that $\|\mathbf{\Phi } \mathbf{G} \mathbf{W}_c\|^2_F = \sum_{p\in{g,b}}\mathrm{tr}(\mathbf{S}_p\mathbf{U})$, $\|\mathbf{h}_{rk}^H\mathbf{\Phi } \|^2 = \mathrm{tr}(\mathbf{\Omega}_k\mathbf{U})$, and $\|\mathbf{\Phi}\|_F^2 = \mathrm{tr}(\mathbf{\Pi}\mathbf{U}) $.
Also, we introduce  $\mathbf{\bar{C}}_{k}=[\mathrm{diag}({{\mathbf{h}}_{{rk}}^H})
{\mathbf{G}} ;\mathbf{h}_{ak}^H ]$ and $\mathbf{{C}}_{k}=\mathbf{\bar{C}}_{k}\mathbf{\bar{C}}_{k}^H$ to transform constraint $\|\mathbf{g}_b\|^2 \ge \|\mathbf{g}_g\|^2$ into $\mathrm{tr}(\mathbf{C}_{b}\mathbf{U})\ge \mathrm{tr}(\mathbf{{C}}_{g}\mathbf{U})$.
After taking the ratio term $\gamma_{b, s_b}$ out of the $\mathrm{log}(\cdot)$ function, we can formulate the reflection beamforming optimization problem as
\begin{subequations}
\begin{align}
~~\!\!\!\!\!\!\!\!({\text{P}}4):~&\mathop {\max }\limits_{{\mathbf{U }}} ~\frac{\mathrm{tr}(\mathbf{\Lambda}_{b,b}\mathbf{U})}
{\mathrm{tr}(\mathbf{\Omega}_b\mathbf{U})\sigma_r^2
+\sigma_b^2}& \label{P4}\\
\text{s.t.}~~~~
&  \mathrm{tr}(\mathbf{{C}}_{b}\mathbf{U})\ge \mathrm{tr}(\mathbf{{C}}_{g}\mathbf{U}),&\label{P4_b}\\
&  \mathrm{tr}(\mathbf{\Lambda}_{g,g}\mathbf{U}) \! \ge \! \gamma_\mathrm{th}\!\Big(\mathrm{tr}(\mathbf{\Lambda}_{g,b}\mathbf{U}) \!+\!\mathrm{tr}(\mathbf{\Omega}_g \mathbf{U})\sigma_r^2 \!+\!\sigma_g^2\Big),&\label{P4_c}\\
&  \sum_{p\in \{g,b\}}\mathrm{tr}(\mathbf{S}_p\mathbf{U})\!+\!
\mathrm{tr}(\mathbf{\Pi} \mathbf{U})\sigma_r^2 \le P_r^\mathrm{max},&\label{P4_d}\\ &\mathrm{tr}(\mathbf{\Lambda}_{w,b}\mathbf{U})\!+\!(1\!-\!\kappa)
\Big(\mathrm{tr}(\mathbf{\Lambda}_{w,g} \mathbf{U})+ \mathrm{tr}(\mathbf{\Omega}_w \mathbf{U})\sigma_r^2\Big)& \nonumber\\
& ~~~~~~~~~~~~~~~~~~~~~~~~\le (\kappa-1) \sigma_w^2 ,&\label{P4_e}\\
&|\mathbf{U}_{n,n}| \le \eta_n^2, ~\forall n,~|\mathbf{U}_{N+1,N+1}|=1,&\label{P4_f}\\
&\mathbf{U} \succeq 0,&\label{P4_g}\\
&\mathrm{rank}(\mathbf{U}) =1.&\label{P4_h}
\end{align}
\end{subequations}
To recast the resulting non-concave fractional programming (FP) problem with rank-one constraint to a convex form, we propose a penalized Dinkelbach approach.
Specifically, we add a penalty-term $\frac{1}{\iota_2 }( {{{\| {\bf{U}} \|}_*} + {{\widehat {\bf{U}}}^{(t)}}} )$ to the objection function's denominator to address the non-convexity of rank-one constraint generated by variable change in \eqref{P4_h}. Here, $\iota_2$ is a control factor and $\widehat {\bf{U}}^{(t)}$ is the upper bound on $-\|{\bf{U}} \|_2$ with closed form $\widehat {\bf{U}}^{(t)} = -\|{\bf{U}} \|_2^{(t)} - {\rm tr}\big({\bf u}_{\max }^{(t)}{({\bf u}_{\max }^{(t)})^H}\big({\bf U} - {{\bf U}^{(t)}}\big)\big) $, where ${\bf u}_{\max }^{(t)}$ represents the eigenvector corresponding to the largest eigenvalue of ${\bf U}^{(t)}$ in the $t$th solution. As a result, the new objective function can be expressed as
\begin{equation}
\frac{{{\rm{tr}}({{\bf{\Lambda }}_{b,b}}{\bf{U}})}}{ {\rm{tr}}({{\bf{\Omega }}_b}{\bf{U}})\sigma _r^2 \!+\! \sigma _b^2+ \frac{1}{\iota_2 }\Big( {{{\| {\bf{U}} \|}_*} \!+\! {{\widehat {\bf{U}}}^{(t)}}} \Big) }. \label{Objective_U}
\end{equation}
By maximizing this fractional objective function, we ensure that ${\rm rank}({\bf U}) = 1$ as $\iota_2 \to 0$ becasue ${\| {\bf{U}} \|}_* -\|{\bf{U}} \|_2 = 0$ is equivalent to ${\rm rank}({\bf U}) = 1$.
Since the above concave-convex single-ratio objective in \eqref{Objective_U} still hinders a direct solution, we further utilize the Dinkelbach transformation \cite{Dinkelbach} to convert it into a concave form. Specifically, let 
\begin{eqnarray}
    f_1({\bf U}) = {{{\rm{tr}}({{\bf{\Lambda }}_{b,b}}{\bf{U}})}}
\end{eqnarray}
and 
\begin{eqnarray}
    f_2({\bf U}) = {\rm{tr}}({{\bf{\Omega }}_b}{\bf{U}})\sigma _r^2  + \sigma _b^2 + \tfrac{1}{\iota_2 }\big( {{{\| {\bf{U}} \|}_*} + {{\widehat {\bf{U}}}^{(t)}}} \big),
\end{eqnarray}
the transformed optimization problem can be expressed as
\begin{subequations}
\begin{align}
({\text{P}}4.1):&\ \mathop {\max }\limits_{{\bf U}} \
  f_1({\bf U})-   u_1   f_2 ({\bf U})   &
\label{P4.1_a}\\
& ~~~{\text{s.t.}}~\eqref{P4_b} -\eqref{P4_f},&\label{P4.1_b} \\
&~~~{\bf U} \succeq  0,&\label{P4.1_c}
\end{align}
\end{subequations}
where the auxiliary variable $u_1$ is updated by $ u_1^{(\ell+1)} = f_1^{(\ell)}({\bf U})/f_2^{(\ell)}({\bf U})$ in the $\ell$th iteration.
Now, problem (P4.1) is concave and $\mathbf{U}$ can be optimized via the existed CVX solver.

The proposed AO algorithm is summarized in Algorithm 1.
To prevent the optimization falling into local optima caused by the zero penalty-term during initialization, we set $u^{(0)} > 0$ in the proposed penalized Dinkelbach approach.
The overall computational complexity of Algorithm 1 is characterized by $\mathcal{O}\left[I_\mathrm{A}(I_{\mathrm{P}_1}(2M)^{3.5}+
I_{\mathrm{P}_2} I_\mathrm{D}(N+1)^{3.5})\right]$, where $I_\mathrm{A}$ and $I_\mathrm{D}$ denote the iteration numbers of the AO loop and Dinkelbach loop, respectively, and $I_{\mathrm{P}_1}$ and $I_{\mathrm{P}_2}$ denote the iteration numbers of the penalty-terms in problems (P3.1) and (P4.1), respectively.

\begin{algorithm}[t]
{
\caption{AO Algorithm for Solving Problem (P1) }
\begin{algorithmic}[1]
\State \small Initialize $s \leftarrow 0$, ${\bf U}^{(0)}$ and ${\bf W}_c^{(0)}$
\State \textbf{repeat} $s \leftarrow s+1$
     \State \indent $t \leftarrow 0$
    \State \indent\textbf{repeat} $t \leftarrow t+1$
    \State \indent\indent  Update ${\bf W}_c^{(t)}$ by solving (P3.1)

    \State \indent \indent $\iota_1  \leftarrow c_1 \iota_1 $
     \State \indent\textbf{until} $ \frac{1}{\iota_1}\!\! \sum_{p \in \{ g,b\} } (\|\mathbf{W}_p\|_*+\widehat{\mathbf{W}}_p^{(t)}) \le \xi_2$
     \State \indent  ${{\bf{W}}_c^{( s )}} \leftarrow {{\bf{W}}_c^{( t )}}$
     \State \indent $t \leftarrow 0$
    \State \indent\textbf{repeat} $t\leftarrow t+1$
        \State \indent\indent Initialize $u_1^{(0)} > 0$,  $\ell \leftarrow 0$
        \State \indent\indent\textbf{repeat}  $\ell \leftarrow \ell+1$
        \State \indent\indent\indent  Update ${\bf U}^{(\ell)}$ by solving (P4.1)
        \State \indent\indent\indent  Update $u_1^{(\ell+1)} = f_1^{(\ell)}({\bf U})/f_2^{(\ell)}({\bf U})$
         \State \indent\indent\textbf{until} $f_1^{(\ell)}({\bf U})-   u_1^{(\ell)} f_2^{(\ell)}({\bf U}) \le 0$
        \State \indent \indent  ${{\bf{U}}^{( t )}} \leftarrow {{\bf{U}}^{( \ell )}}$, $\iota_2  \leftarrow c_2 \iota_2 $
    \State \indent\textbf{until} $\frac{1}{\iota_2 }( {{{\| {\bf{U}} \|}_*^{(t)}} + {{\widehat {\bf{U}}}^{(t)}}} ) \le \xi_3$
    \State \indent   ${{\bf{U}}^{( s  )}} \leftarrow {{\bf{U}}^{( t )}}$
 \State \textbf{until}  $R_{b,s_b}^{(s)}- R_{b,s_b}^{(s-1)} \le \xi_1 $
\end{algorithmic}
}
\end{algorithm}
\subsection{Proposed Solution to Problem (P2)}

To obtain the solution to the non-concave optimization problem (P2), we also transform problem (P2) into two sub-problems of optimizing the transmission beamforming $\mathbf{W}$ and reflection beamforming $\mathbf{\Phi}$, respectively. The resultant sub-problems are efficient solved by utilizing an AO Algorithm similar to Algorithm 1.

\emph{1) Transmission Beamforming Optimization:} For any given feasible $\mathbf{\Phi}$, problem (P2) can be reduced to optimize ${\bf W}$ only. We introduce the auxiliary variable $\tilde{\mathbf{{ W}}}_s= \mathbf{W}_s\mathbf{W}_s^H$ and obtain ${\left\| {\mathbf{g}_k^H {{{\mathbf{W}}_{s}}} } \right\|^2}= \mathrm{tr}(\mathbf{\Upsilon}_k\tilde{\mathbf{ W}}_s)$. Thus, the optimization problem with respect to ${\bf W}$ can be formulated as
\begin{subequations}
\begin{align}
~~~\!\!\!\!\!\!\!\!({\text{P}}5):& \mathop {\max }\limits_{ \mathbf{W}_g, \mathbf{W}_b,\tilde{\mathbf{{W}}}_s, \mathbf{J}}
\mathrm{tr}(\mathbf{\Upsilon}_b\mathbf{W}_b)&
\label{P5_a}\\
 {\text{s.t.}}~~&\mathrm{tr}(\bar {\mathbf{ W}}) \le P_a^{\max },&\label{P5_b}\\
 &\mathrm{tr}(\mathbf{W}_g )\ge \mathrm{tr}(\mathbf{W}_b ),&
\label{P5_c}\\
\!\!\!\!\!& \mathrm{tr}(\mathbf{\Upsilon}_g \mathbf{W}_g) \ge \gamma_\mathrm{th} \Big( \mathrm{tr}\big(\mathbf{\Upsilon}_g \mathbf{W}_b\big)+\mathrm{tr}(\mathbf{H}_{rg}
\mathbf{\Psi}) \sigma _r^2 \!+\! \sigma _{g}^2 \Big), &\label{P5_d}\\
 \!\!\!\!\!&  \mathrm{tr}( \mathbf{J}) \le \mu,~ \bar{\mathbf{ F}}(\bar{\mathbf{W}}) -\mathbf{J}\succeq 0&
 \label{P5_e}\\
&\mathrm{tr}(\mathbf{\Gamma} \bar{\mathbf{{W}}})+ \mathrm{tr}(\mathbf{\Psi})\sigma_r^2 \le P_r^\mathrm{max},&
\label{P5_f}\\
&\mathrm{tr}(\mathbf{\Upsilon}_w\mathbf{W}_b)\!+\!(1\!-\!\kappa)
\Big(\mathrm{tr}\big(\mathbf{\Upsilon}_w (\mathbf{W}_g \!+\!\tilde{\mathbf{{W}}}_s ) \big)& \nonumber\\
 &~~~~~~~~~~~~~~~~~~\!+\!\mathrm{tr}(\mathbf{H}_{rw}\mathbf{\Psi})\sigma_r^2\Big) \le (\kappa\!-\!1)\sigma_w^2 ,&\label{P5_g}\\
 & \mathbf{W}_g \succeq 0, ~\mathbf{W}_b \succeq 0, ~ \tilde{\mathbf{{W}}}_s \succeq 0, ~ \mathbf{J} \succeq 0 & \label{P5_h}\\
& \mathrm{rank}(\mathbf{W}_g) =1,~\mathrm{rank}(\mathbf{W}_b) =1,&\label{P5_i}
\end{align}
\end{subequations}
where $\bar{\mathbf{W}} = \mathbf{W}_g+\mathbf{W}_b+ \tilde{\mathbf{W}}_s$. Then, we apply the penalty-term method same as in problem (P3.1) to cope with the non-convexity of rank-one constraint in problem (P5), yielding the following transformed optimization problem:
\begin{subequations}
\begin{align}
~~~\!\!\!\!\!({\text{P}}5.1):& \mathop {\max }\limits_{ \mathbf{W}_g, \mathbf{W}_b,\tilde{\mathbf{{W}}}_s,\mathbf{J} }
\mathrm{tr}(\mathbf{\Upsilon}_b\mathbf{W}_b) &
\nonumber \\
& \qquad \qquad - \frac{1}{\iota_3} \sum\limits_{p \in \{ g,b\} } \Big(\|\mathbf{W}_p\|_*+\widehat{\mathbf{W}}_p^{(k)}\Big)&
\label{P3.1_a}\\
 &~\qquad{\text{s.t.}}~\eqref{P5_b}-\eqref{P5_h}.&
\label{P3.1_b}
\end{align}
\end{subequations}
Now, the above transmission beamforming problem is a concave optimization problem.

\emph{2) Reflection Beamforming Optimization:}
For any given $ \mathbf{W}$, problem (P2) can be reduced to optimize $\mathbf{\Phi}$ only.
With the introduced auxiliary variables in the previous section, we separately define $\mathbf{\bar{\Lambda}}_{k,j}=[\mathrm{diag}({{\mathbf{h}}_{{rk}}^H})
{\mathbf{G}} \mathbf{w}_j;\mathbf{h}_{ak}^H \mathbf{w}_j]$ with $\mathbf{w}_j$ bing the $j$th, $j\in \{ 1,...,2+M\}$, column of the beamforming matrix $\mathbf{W}$ and $\mathbf{S}_j= \mathrm{diag}\left( |[\mathbf{s}_j]_1|^2,...,
|[\mathbf{s}_j]_N|^2, 0 \right)$ with $\mathbf{s}_j=
\mathbf{G}\mathbf{w}_j$, and construct
$\mathbf{\Lambda}_{k,j}= \mathbf{\bar{\Lambda} }_{k,j} \mathbf{\bar{\Lambda}}_{k,j}^H$, yielding the expressions ${{{\left| {\mathbf{g}_k^H{{\mathbf{w}}_{j}}} \right|}^2}}=\mathrm{tr}(\mathbf{\Lambda}_{k,j}\mathbf{U})$ and $\|\mathbf{\Phi } \mathbf{G} \mathbf{W}\|^2_F = \sum_{j=1}^{2+M}\mathrm{tr}(\mathbf{S}_j\mathbf{U})$.
As a result, the reflection beamforming optimization problem is formulated as
\begin{subequations}
\begin{align}
~~\!\!\!\!\!\!\!\!({\text{P}}6):~&\mathop {\max }\limits_{{\mathbf{U }}} ~\frac{\mathrm{tr}(\mathbf{\Lambda}_{b,2}\mathbf{U})}
{\mathrm{tr}(\mathbf{\Omega}_b\mathbf{U})\sigma_r^2
+\sigma_b^2}& \label{P6}\\
\text{s.t.}~~~~
&  \mathrm{tr}(\mathbf{{C}}_{b}\mathbf{U})\ge \mathrm{tr}(\mathbf{{C}}_{g}\mathbf{U}),&\label{P6_b}\\
&  \mathrm{tr}(\mathbf{\Lambda}_{g,1}\mathbf{U}) \! \ge \! \gamma_\mathrm{th}\!\Big(\mathrm{tr}(\mathbf{\Lambda}_{g,2}\mathbf{U}) \!+\!\mathrm{tr}(\mathbf{\Omega}_g \mathbf{U})\sigma_r^2 \!+\!\sigma_g^2\Big),&\label{P6_c}\\
& \sum_{j=1}^{2+M}\mathrm{tr}(\mathbf{S}_j\mathbf{U})\!+\!
\mathrm{tr}(\mathbf{\Pi} \mathbf{U})\sigma_r^2 \le P_r^\mathrm{max} ,&\label{P6_d}\\ &\mathrm{tr}(\mathbf{\Lambda}_{w,2}\mathbf{U})\!+\!(1\!-\!\kappa)\!
\Bigg(\! \sum_{j=1,j\neq2}^{2+M}\!\!\!\!\!\mathrm{tr}(\mathbf{\Lambda}_{w,j} \mathbf{U})\!+\! \mathrm{tr}(\mathbf{\Omega}_w \mathbf{U})\sigma_r^2\!\!\Bigg)& \nonumber\\
& ~~~~~~~~~~~~~~~~~~~~~~~~\le (\kappa-1) \sigma_w^2 ,&\label{P6_e}\\
&|\mathbf{U}_{n,n}| \le \eta_n^2, ~\forall n,~|\mathbf{U}_{N+1,N+1}|=1,&\label{P6_f}\\
&\mathbf{U} \succeq 0.&\label{P6_g}\\
&\mathrm{rank}(\mathbf{U}) =1.&\label{P6_h}
\end{align}
\end{subequations}
Similar to the approach adopted in problem (P4), we transform problem (P6) into a concave one by applying the penalized Dinkelbach approach, which is formulated as
\begin{subequations}
\begin{align}
({\text{P}}6.1):&\ \mathop {\max }\limits_{{\bf U}} ~
  f_3 ({\bf U})-   u_2   f_4 ({\bf U})   &
\label{P4.1_a}\\
&~~~{\text{s.t.}} ~\eqref{P6_b} -\eqref{P6_g},&\label{P6.1_b}
\end{align}
\end{subequations}
where 
\begin{eqnarray}
    f_3({\bf U}) = {{{\rm{tr}}({{\bf{\Lambda }}_{b,2}}{\bf{U}})}}, 
\end{eqnarray}
\begin{eqnarray}
    f_4({\bf U}) = {\rm{tr}}({{\bf{\Omega }}_b}{\bf{U}})\sigma _r^2  + \sigma _b^2 + \tfrac{1}{\iota_4 }\big( {{{\| {\bf{U}} \|}_*} + {{\widehat {\bf{U}}}^{(t)}}} \big)
\end{eqnarray}
with $\iota_4$ being a penalty factor, and $u_2$ is a auxiliary variable updated by $ u_2^{(\ell+1)} = f_3^{(\ell)}({\bf U})/f_4^{(\ell)}({\bf U})$ in the $\ell$th iteration. After the above transformation, problem (P6.1) can be efficiently solved by the existing convex solvers. Similar to Algorithm 1, problem (P2) can be efficiently solved by iteratively solving the decoupled sub-problems (P5.1) and (P6.1)) with an AO Algorithm. Since the algorithm structure for solving problem (P2) is the same as Algorithm 1, we omit the details of the algorithm for solving problem (P2). In addition, the overall computational complexity of the algorithm for solving problem (P2) can be characterized by $\mathcal{O}\left[I_\mathrm{A}(I_{\mathrm{P}_3}(3M)^{3.5}+
I_{\mathrm{P}_4} I_\mathrm{D}(N+1)^{3.5})\right]$, where $I_{\mathrm{P}_3}$ and $I_{\mathrm{P}_4}$ denote the iteration numbers of the penalty-terms in problems (P5.1) and (P6.1), respectively.

 \begin{figure}[t]
    \begin{center}
    \includegraphics[width=3.1in]{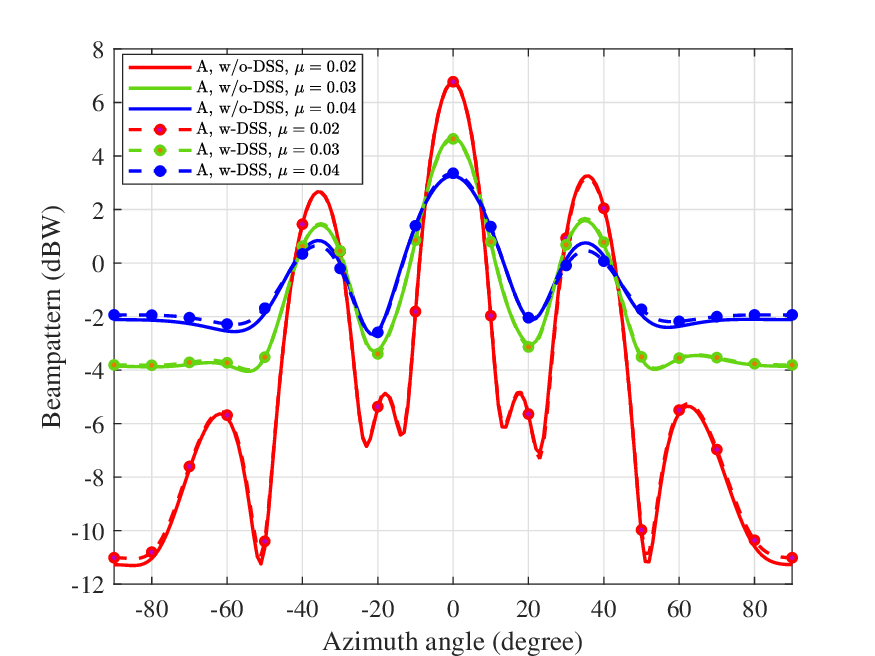}
    \caption{Impacts of CRB threshold on beampattern}
    \label{fig:beampattern}
    \end{center}
    \vspace{-0.25in}
\end{figure}

\section{Numerical Results}

In this section, numerical results are provided to validate the system performance of the active-RIS-aided NOMA-ISAC system. Unless otherwise specified, we set $M = 8$, $N=16$, $L=1024$, $P^\mathrm{max}_a=30$ dBm, $P^\mathrm{max}_r=25$ dBm, $\sigma^2_b=\sigma^2_g=\sigma^2_w=\sigma^2_r=-90$ dBm, $R_g^\mathrm{min}=1$ bps/Hz, $c_1=c_2=\xi_1=10^{-2}$, $\xi_2=\xi_2= 10^{-4}$, and $\eta_n^2=\eta^2$, $\forall n$. In the simulations, Alice, Bob, Grace, Willie, and active-RIS are located at $(0,0)$ m, $(80,10)$ m, $(90,0)$ m, $(70,5)$ m, and $(75,30)$ m, respectively, in an x-y coordinate plane. There are $Q=3$ moving targets located at $\hat{\theta}_1=-35^\circ$, $\hat{\theta}_2=0^\circ$, and $\hat{\theta}_3=35^\circ$, respectively, related to x-axis, and have distances of $40$ m, $50$ m, and $35$ m from Alice, respectively. The velocities of the three targets are  $6$ m$/$s, $14$ m$/$s, and $10$ m$/$s, respectively, and their RCSs  satisfy $\{\sigma_q^2\}_{q=1}^Q = 1$. For the channel fading, the path-loss is modeled by $\mathcal{L}=\mathcal{L}_0 d^{-\chi}$ with $\mathcal{L}_0=-30$ dB and $d$ denoting the distance between the two terminals. Specifically, we set $\chi=3.5$ for the Alice-$k$ link, $\chi=2.3$ for Alice-$q$ link, and $\chi=2.2$ for the channels associated with the active-RIS \cite{ARIS_secure}. The Rician factor for the active-RIS-associated links is set as $\beta=3$ dB, while for the Alice-$k$ link is set as $\beta=0$. In the following discussions, we compare the covert communication performance of the proposed active-RIS-aided NOMA-ISAC scheme against the passive-RIS-aided and without-RIS counterparts. To simulate the results of passive-RIS-aided NOMA-ISAC scheme, the covert rate maximization problems using both the w/o-DSS and w-DSS schemes are formulated by modifying the corresponding objectives and constraints. Specifically, constraint \eqref{P1_1f} is modified as $| {\bf \Phi}_{{n,n}}  | = 1 $,  constraint  \eqref{RIS_power_constraint} is omitted, and $\sigma_r^2$ is set to be zero in the corresponding objectives and constraints. In the following figures, the performance curves achieved by the active-RIS-aided and passive-RIS-aided NOMA-ISAC schemes are denoted as “A” and “P”, respectively, while the performance curves by the benchmark scheme without-RIS scheme are denoted as “w/o-RIS” in the legends.

The optimaml transit beampatterns achieved by the w-DSS and w/o-DSS schemes under various CRB thresholds are plotted in Fig \ref{fig:beampattern}, where we set $\eta= 40$ and $\varepsilon=0.1$. Specifically, the transmit beamforming matrices of the w/o-DDS and w-DSS schemes are the optimized solutions of problems (P3.1) and (P5.1), respectively.
The results in Fig. \ref{fig:beampattern} show that the three moving targets are accurately captured in the mainlobes of the transmit beampatterns by the w-DSS and w/o-DSS schemes. Besides, the results in Fig. \ref{fig:beampattern} reveals the impacts of CRB threshold on the transmit beampattern, i.e., a smaller CRB threshold can achieve a better beampattern because a smaller CRB threshold requires more transmit power to be allocated towards the target directions. Furthermore, since the CRB threshold constraints are the same for the w-DSS and w/o-DSS schemes, the transmit beampatterns achieved by the two schemes are almost identical in the mainlobes. Therefore, both the w-DSS and w/o-DSS schemes satisfy the sensing performance requirements in the considered system.  

\begin{figure}[t]
    \begin{center}
    \includegraphics[width=3.1in]{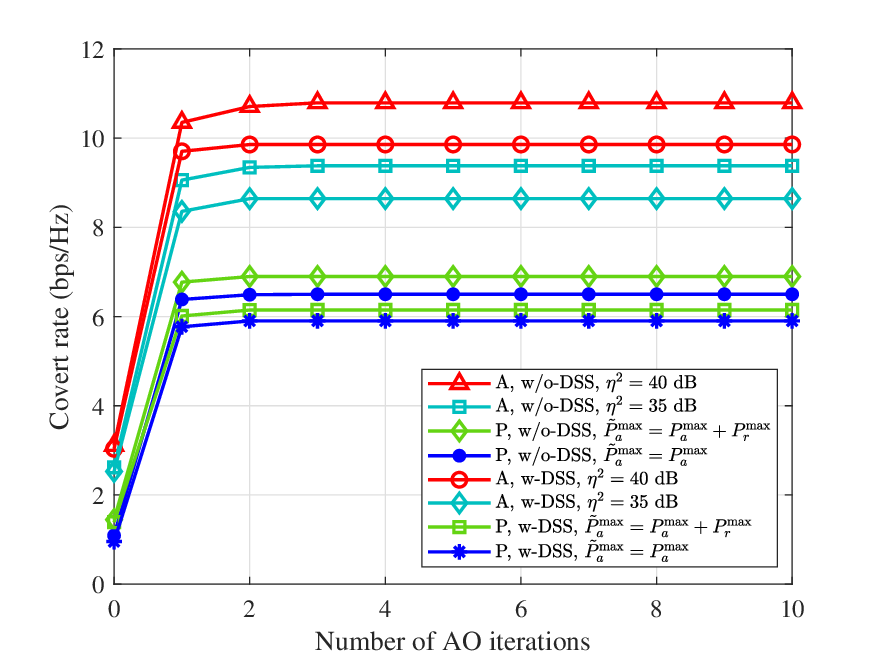}
    \caption{Convergence performance of AO algorithm}
    \label{fig:AO_Convergence}
    \end{center}
    \vspace{-0.25in}
\end{figure}

We show the convergence performance of the proposed AO algorithm in Fig. \ref{fig:AO_Convergence}, where $\varepsilon = 0.1$ and $\mu = 0.08$. The covert rates versus the number of AO iterations under different settings of the active-RIS-aided and passive-RIS-aided systems are presented to verify the convergence of our proposed AO algorithm. Considering that the active-RIS introduces additional power supplies, the total system resource consumption is higher than the passive-RIS-aided system. For a relatively fair comparison with the passive-RIS-aided system, we consider two power budgets for the passive-RIS-aided system in which Alice has a power budget of either $\tilde{P}^\mathrm{max}_a= P^\mathrm{max}_a$ or $\tilde{P}^\mathrm{max}_a= P^\mathrm{max}_a + P^\mathrm{max}_r$. The curves of Fig. \ref{fig:AO_Convergence} show that the proposed AO algorithm converges within $10$ iterations for all the considered system settings. Furthermore, the covert rate almost no longer increases after $3$ iterations, which verifies the fast convergence and effectiveness of the proposed AO algorithm. Since the covert rate achieved in each iteration of transmission and reflection beamforming optimization consistently demonstrates a non-decreasing pattern, a limited number of iterations is sufficient for achieving convergence in the proposed AO algorithm.

\begin{figure}[t]
    \begin{center}
    \includegraphics[width=3.4in]{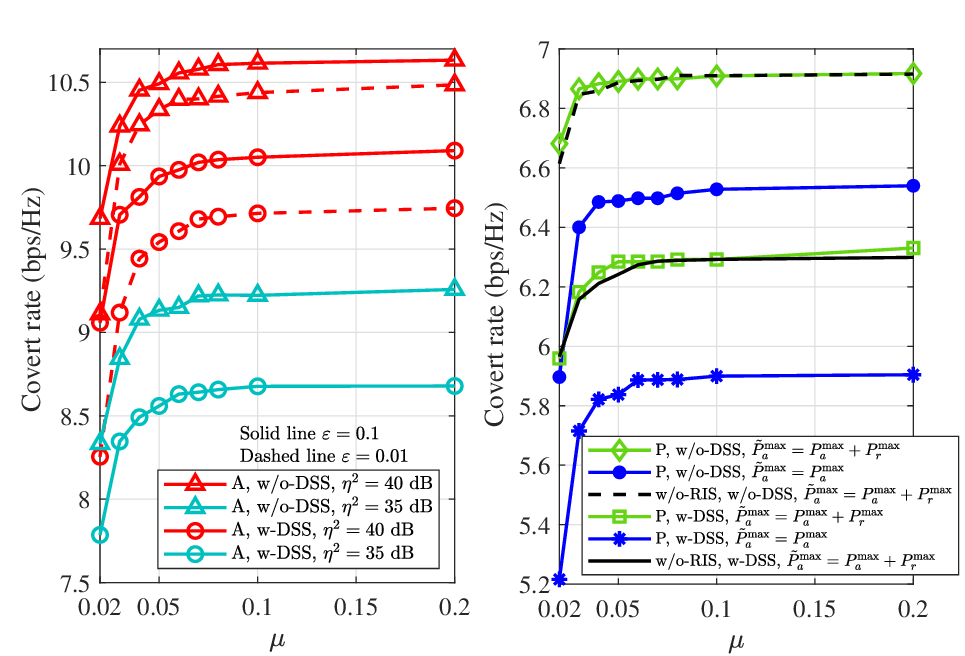}
    \caption{Covert rate versus CRB threshold with $Q=3$.}
    \label{fig:epsilon}
    \end{center}
    \vspace{-0.25in}
\end{figure}

In Fig. \ref{fig:epsilon}, we plot the curves of the covert rates achieved by the w/o-DSS and w-DSS schemes versus the CRB threshold to investigate the trade-off between covert communication performance and target parameter estimation performance.
The without-RIS system is considered to have power budget $\tilde{P}^\mathrm{max}_a= P^\mathrm{max}_a + P^\mathrm{max}_r$ for fair comparison. 
The curves in Fig. \ref{fig:epsilon} clearly show the trade-off between the covert communication performance and target parameter estimation performance. As the CRB threshold relaxes, the priority is shifted to covert transmission, resulting in the increased covert rates for all the considered schemes. However, in the high CRB threshold regime, the covert rates achieved by all the schemes converge to be constant. Furthermore, both a higher $\eta^2$ and a looser covertness constraint lead to a higher covert rate, as verified by Fig. \ref{fig:epsilon}(a).
When the w/o-DSS scheme is applied, the active-RIS-aided and passive-RIS-aided systems, as well as the w/o-RIS system achieve a better trade-off compared to the w-DSS scheme. The reason for this phenomenon is that the introduction of the DSS inevitably results in additional power consumption at Alice to realize sensing, thus leading to a degradation of the covert communication performance. The same phenomenon occurs at a stricter covertness level, i.e., $\varepsilon$ decreases from 
$0.1$  to $0.01$, and smaller reflection amplitude of active-RIS elements, i.e., $\eta^2 = 35 $ dB,  as shown in Fig. \ref{fig:epsilon}(a). In addition, the curves in Fig. \ref{fig:epsilon} also reveal that the active-RIS-aided system has a significant advantage in the trade-off performance compared to the passive-RIS-aided system and w/o-RIS system. Furthermore, the passive-RIS provides negligible covert rate gain compared to w/o-RIS system. 
Due to the additional reflection gain loss and available covert transmission power loss, the passive-RIS-aided system using the w-DSS scheme with $\tilde{P}^\mathrm{max}_a= P^\mathrm{max}_a$ achieves the poorest trade-off performance. 

\begin{figure}[t]
    \begin{center}
    \includegraphics[width=3.1in]{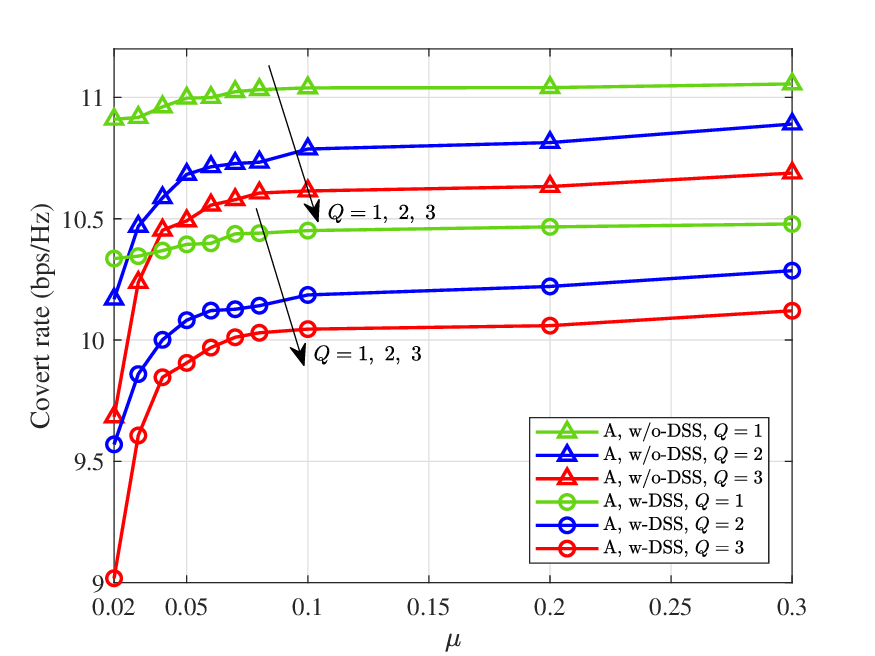}
    \caption{Covert rate versus CRB threshold with $Q=1, 2, 3$.}
    \label{fig:variousQ}
    \end{center}
    \vspace{-0.25in}
\end{figure}

\begin{figure}[t]
    \begin{center}
    \includegraphics[width=3.1in]{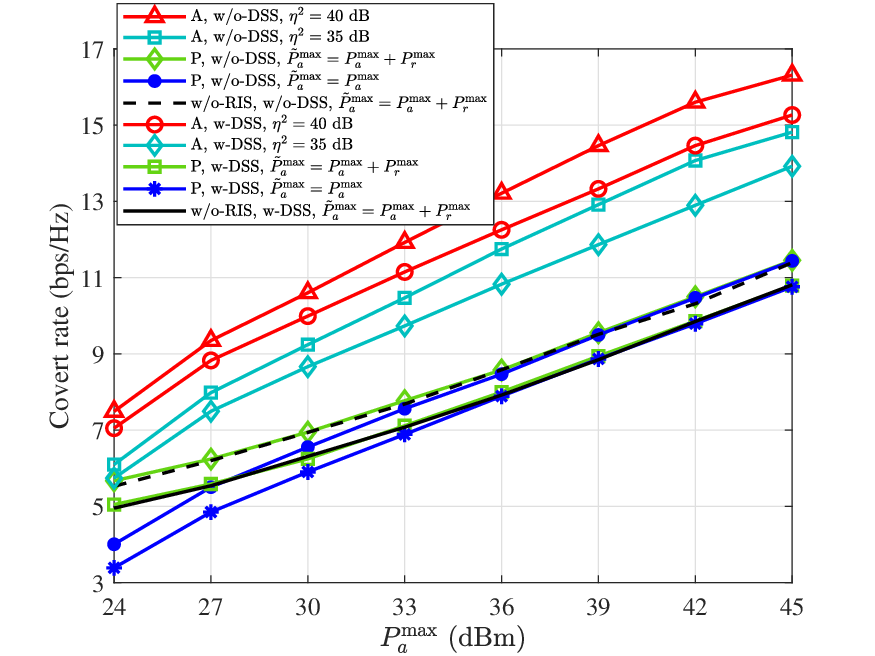}
    \caption{Covert rate versus transmit power.}
    \label{fig:Pa}
    \end{center}
    \vspace{-0.25in}
\end{figure}

In Fig. \ref{fig:variousQ}, to obtain more insights of the proposed schemes, we investigate the impacts of the moving target number on the trade-off between covert communication and sensing performances under the setting of $\varepsilon=0.1$ and $\eta = 40$ dB. When $Q=1$, Alice senses the target located at $\hat{\theta}_1$. When $Q=2$, we assume the targets located at $\hat{\theta}_1$ and $\hat{\theta}_2$ need to be detected, while all the three targets are considered for parameter estimation when $Q=3$. 
Analogous to the results in Fig. \ref{fig:epsilon}, the covert rates achieved by the cases of $Q=1$ and $Q=2$ increase with the increase of CRB threshold, i.e. the priority shifts to covert transmission. Furthermore, we observe that the w/o-DSS scheme also exhibits an explicit trade-off region gain over the w-DSS scheme when $Q=1$ and $Q=2$. Attributing to the additional available transmit power for information transmission, the w/o-DSS scheme achieves a superior covert rate than the w-DSS scheme at the rightmost corner point for the considered three cases of target number. Besides, as the number of targets increases, the trade-off regions of both the w/o-DSS scheme and w-DSS scheme become worse due to the loss of the transmission beamforming power at each individual target.

In Fig. \ref{fig:Pa}, we investigate the impacts of the transmit power budget on the covert rate for different schemes, where we set $P^\mathrm{max}_r=25$ dBm and $\mu = 0.08$. From Fig. \ref{fig:Pa}, we can see that the covert rates achieved by all the schemes increase with the increasing $P_a^\mathrm{max}$. 
Furthermore, the active-RIS-aided systems always outperform the passive-RIS-aided systems and w/o-RIS systems no matter when which transmission scheme is adopted. As expected, the improvement in covert performance is very little for the passive-RIS-aided systems compared to the w/o-RIS system due to the “multiplicative fading effect”.
The curves in Fig. \ref{fig:Pa} confirm that a higher $\eta^2$ leads to a higher covert rate, which indicates that the constraints \eqref{RIS_power_constraint} and \eqref{RIS_power_constraint1} are inactive, respectively. Thanks to the additional covert rate gain provided by the active-RIS and more available transmit power for covert transmission in the absence of the DSS, the active-RIS system using the w/o-DSS scheme achieves a considerably large covert rate in the high transmit power regime. The curves in Fig. \ref{fig:Pa} reveal that the passive-RIS-aided system with $\tilde{P}^\mathrm{max}_a= P^\mathrm{max}_a + P^\mathrm{max}_r$ achieves a higher covert rate than the passive-RIS-aided system with $\tilde{P}^\mathrm{max}_a= P^\mathrm{max}_a$ in the low transmit power regime due to the additional transmit power budget, while the achieved covert rate of the consider two passive-RIS-aided systems tend to overlap with the increasing of $P^\mathrm{max}_a$. Thus, in the high $P^\mathrm{max}_a$ regime, transmit power enhancement cannot achieve a higher covert rate in the passive-RIS-aided system.  

\begin{figure}[t]
    \begin{center}
    \includegraphics[width=3.1in]{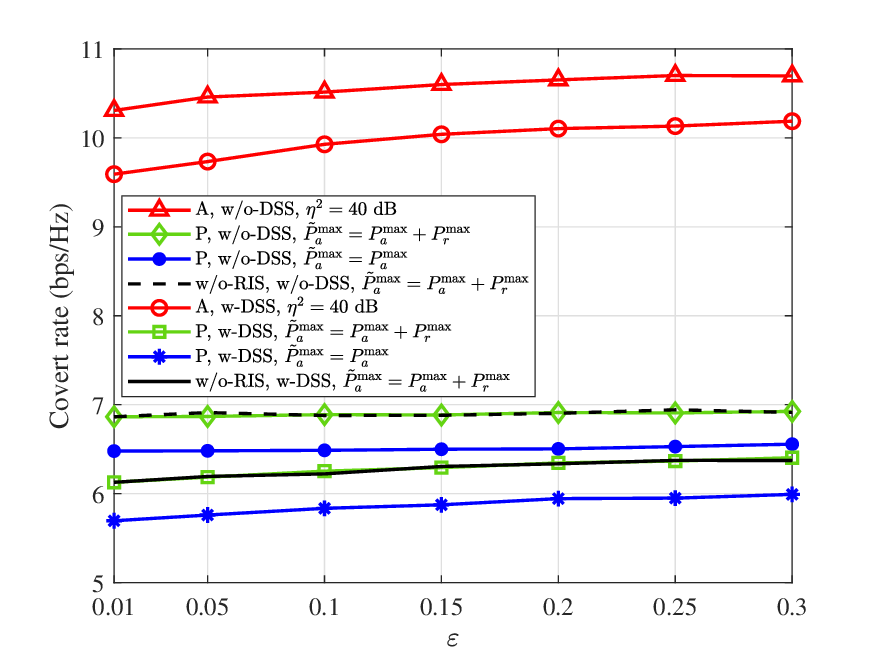}
    \caption{Covert rate versus covertness level $\varepsilon$.}
    \label{fig:varepsilon}
    \end{center}
    \vspace{-0.25in}
\end{figure}
  
Fig. \ref{fig:varepsilon} shows the covertness level $\varepsilon$ versus the covert rate under the two transmission schemes, where we set $\mu = 0.05$. Both the passive-RIS-aided system with two considered power budgets and the w/o-RIS system are taken into consideration. As clearly shown by the curves in Fig. \ref{fig:varepsilon}, the active-RIS-aided system exhibits an improved covertness performance with the increase of $\varepsilon$ compared with the passive-RIS-aided system, which suffers the “multiplicative fading effect”, resulting in negligible covert rate gain compared to the without-RIS system. This result is consistent with the theoretical analysis, i.e., when $\varepsilon$ becomes larger, the covertness constraint becomes loose, and the allocatable transmit power for $\mathbf{w}_b$ and the reflection beamforming power towards Bob is larger, which leads to a larger covert rate for the active-RIS-aided system. Moreover, in the passive-RIS-aided system, as the covertness constraint is relaxed, the increase of the covert rate achieved by the w/o-DSS scheme is slower than that of the w-DSS scheme. This result is because the DSS can also be utilized as the covert medium to conceal the covert transmission from Alice to Bob, resulting in the covert rate growth achieved by the w-DSS scheme being larger than that of the w/o-DSS scheme as the covertness constraint loosing. Again, the active-RIS-aided system with the w/o-DSS scheme achieves a higher covert rate than with the w-DSS scheme, while the same thing occurs in the without-RIS system and passive-RIS-aided system, no matter which transmit power budget is applied.

\begin{figure}[t]
    \begin{center}
    \includegraphics[width=3.1in]{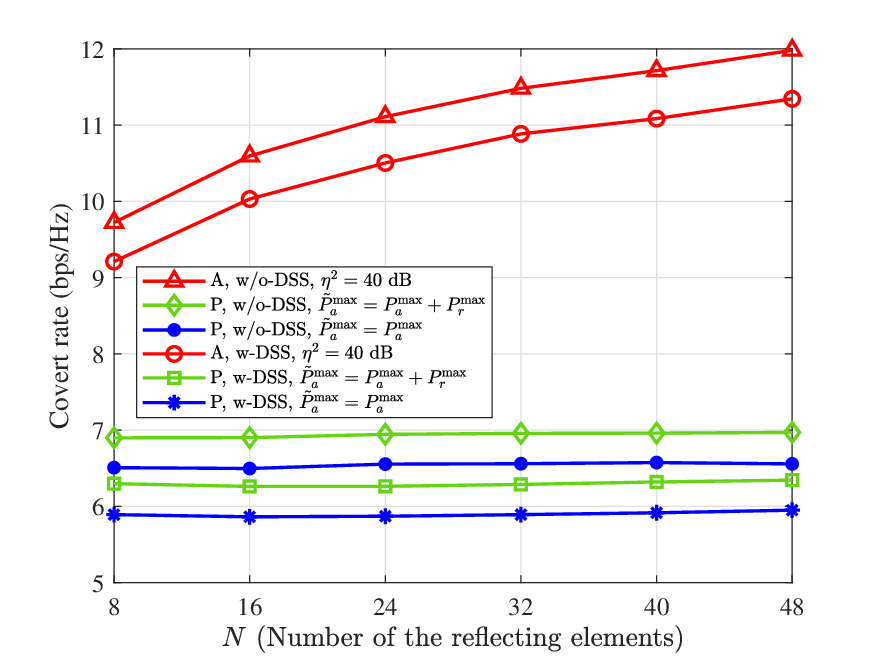}
    \caption{Covert rate versus number of reflecting elements.}
    \label{fig:N}
    \end{center}
    \vspace{-0.25in}
\end{figure}

In the next, the impacts of the number of the RIS reflecting elements, the number of the transmit antennas, and the horizontal position of the RIS on the covert performance are respectively investigated. Unless otherwise stated, we set $\varepsilon=0.1$, and $\mu = 0.08$ in the simulations. The curves in 
Fig. \ref{fig:N} compare  the covert rate performance versus the number of the RIS elements $N$. For the active-RIS-aided system, it is obvious that more reflecting elements of the active-RIS provide more reflection beamforming gain and achieve a higher covert rate due to more DoFs exploited to manipulate the propagation environments. However, the additional reflection beamforming gain is negligible for passive-RIS-aided system, again due to the “multiplicative fading effect”. When the w/o-DSS scheme is applied, both the active-RIS-aided and passive-RIS-aided systems obtain a higher covert rate than the w-DSS scheme. 
Furthermore, the covert rate achieved by the active-RIS-aided system is significantly higher than that of the passive-RIS-aided system, no matter which transmit power budget is applied. The results also show that the passive-RIS-aided system with $\tilde{P}^\mathrm{max}_a= P^\mathrm{max}_a + P^\mathrm{max}_r$ is better than the passive-RIS-aide system  with $\tilde{P}^\mathrm{max}_a= P^\mathrm{max}_a$ due to more transmit power budget at Alice. By comparing the considered two transmission schemes, we see that additional reflecting elements provide more marked covert rate gains in the w/o-DSS scheme provides. To sum up, exploiting the NOMA signal only can achieve a better covert communication performance than utilizing the DSS in the considered RIS-aided NOMA-ISAC systems.

\begin{figure}[t]
    \begin{center}
    \includegraphics[width=3.1in]{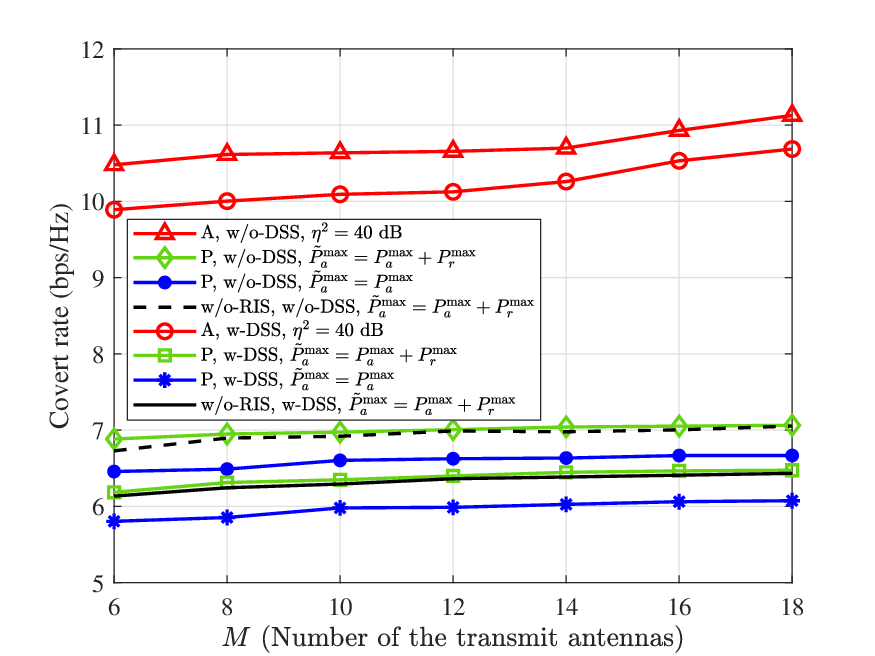}
    \caption{Covert rate versus number of transmit antennas.}
    \label{fig:M}
    \end{center}
    \vspace{-0.25in}
\end{figure}

In Fig. \ref{fig:M}, we investigate the impacts of the number of transmit antennas $M$ on the covert communication performance under different schemes. It can be seen that the covert rates achieved by all the considered schemes increase with the increasing $M$ owing to the enhanced array gain. The covert rate achieved in the active-RIS-aided system increases smoothly at first from $M=6$ to $M=14$, and then increases more dramatically with the continued increasing $M$. 
This result is due to the extra spatial DoFs introduced by the additional transmit antennas, thereby increasing the transmission beamforming resolution. Besides, the covert communication performance of the active-RIS-aided system hugely outperforms the passive-RIS-aided system and w/o-RIS system due to the additional covert rate gain provided by the larger reflecting amplitude of the active-RIS. 
From the perspective of covert communications, although the phase shift modulation of the passive-RIS can be employed to reduce the DEP at Willie, the passive-RIS reflection gain is negligible in the considered NOMA-ISAC system due to the severe "multiplicative fading effect".
For the passive-RIS-aided system, the transmit power budget with $\tilde{P}^\mathrm{max}_a= P^\mathrm{max}_a + P^\mathrm{max}_r$ scheme provides additional SNR for Bob compared to the transmit power budget with $\tilde{P}^\mathrm{max}_a= P^\mathrm{max}_a$ scheme, resulting in a higher covert rate, as indicated in Fig. \ref{fig:M}.
Fig. \ref{fig:M} also reveals that the presence of DSSs degrades the covert rate because the introduction of $\mathbf{s}_s$ inevitably results in additional power consumption at Alice, thus leading to less transmit power for covert transmission.

\begin{figure}[t]
    \begin{center}
    \includegraphics[width=3.1in]{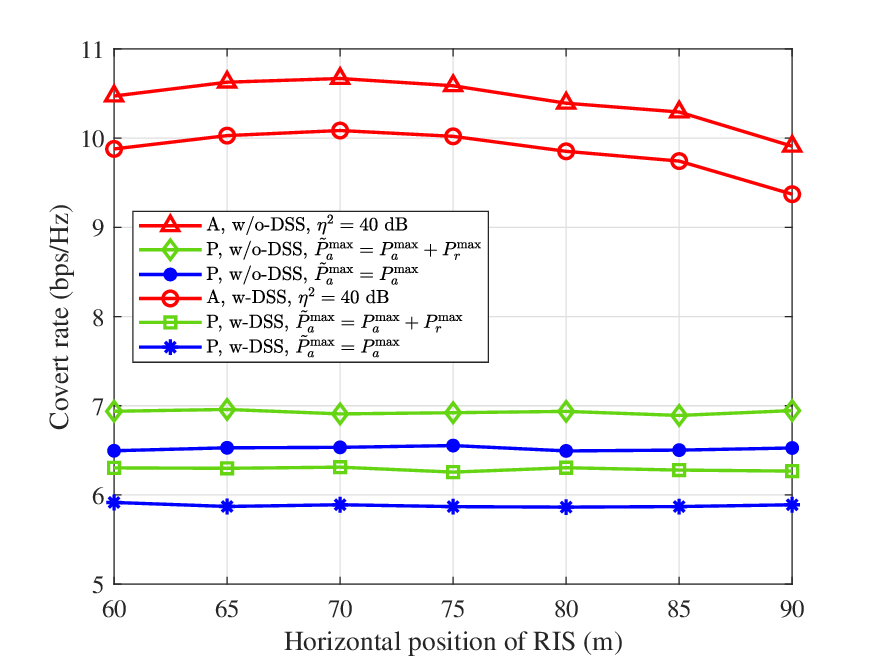}
    \caption{Impacts of RIS location on covert rate.}
    \label{fig:RIS_loc}
    \end{center}
    \vspace{-0.25in}
\end{figure}

Fig. \ref{fig:RIS_loc} shows the covert rate versus the RIS's horizontal position (x-coordinate) under different RIS schemes with the considered two transmission schemes, where the vertical position of RIS is fixed at $30$ m. 
It shows that the covert rates achieved by active-RIS-aided system increase with increasing horizontal position, until reaching about $70$ m and beginning to decrease, which indicates that when the active-RIS is properly deployed close to Bob and Willie, the highest covert rate gain of reflection beamforming can be obtained. 
On the contrary, the covert rates achieved by the passive-RIS-aided system change slightly with the increase of the horizontal position of the RIS, whereas the achieved covert rates are far less than those achieved by the active-RIS-aided system since the reflecting amplitude is limited at $\eta = 1$.
Furthermore, the passive-RIS-aided system with transmit power budget $\tilde{P}^\mathrm{max}_a= P^\mathrm{max}_a + P^\mathrm{max}_r$ scenario achieves a higher covert rate than the passive-RIS-aided system with transmit power budget $\tilde{P}^\mathrm{max}_a= P^\mathrm{max}_a$ scenario, which is expected due to the additional power gain.
Fig. \ref{fig:RIS_loc} also reveals that the covert performance of the w/o-DSS  scheme outperforms that of the w-DSS scheme for both active-RIS-aided and passive-RIS-aided systems. 
Intuitively, when the DDSs are utilized as the shield to cover up the covert transmission for Bob, the covertness level should be better and the lower DEP at Willie may be achieved. 
Actually, for the considered NOAM-ISAC system, the shield of the NOMA public user's signal is strong enough to achieve ideal covert performance at the same covertness level while allowing more transmit power budget to improve the covert rate at the NOMA covert user.
In a nutshell, for the considered NOMA-ISAC system, the DSS is an obstruction to achieve the best covert performance and trade-off between communication and sensing. 

\section{Conclusions}

In this paper, two superposition transmission schemes, namely, the w-DSS and w/o-DSS schemes, were applied to enhance the covert communication and sensing performances for the active-RIS-aided NOMA-ISAC system. To tackle the non-concave covert rate maximization problem, in which the joint optimization of the transmission and reflection beamforming is needed, we decoupled the optimizations of the transmission and reflection beamforming and proposed an AO algorithm to maximize the covert rate. The rank-one constrained fractional programming sub-problems were efficiently solved by introducing the penalized Dinkelbach transformation. The numerical results verified the superiority of the active-RIS-aided NOMA-ISAC system in terms of covert rate and trade-off between communication and sensing performances over that of the passive-RIS-aided and without-RIS NOMA-ISAC system. It was shown that the w/o-DSS scheme achieves a better system performance than the w-DSS scheme. Specifically, it was clarified that exploiting only the NOMA signal achieves the same sensing performance as exploiting the superposition of NOMA signal and DSS, while the additional transmission power gain can be exploited by the w/o-DSS scheme to enhance the covert communication performance.

\section*{Appendix}
\section*{derivation of the FIM in \eqref{FIM}}
The matrix $\mathbf{F}$ is related to four target parameters $\bm{ \xi}_q=[\theta_{aq},\Re\{\alpha_q\},\Im\{\alpha_q\},\mathcal{F}_{D_q}]^T$, $\forall q\in \mathcal{Q}$. According to \eqref{eacho_signal}, the received free-noise signal at Alice is given by $\bm{\eta}[l] = \mathbf{y}_a[l] - \mathbf{z}_a[l] $. The derivatives of $\bm \eta [l]$ with respect to each parameter can be calculated as
\begin{eqnarray}
\frac{\partial \bm \eta[l] }{\partial \mathcal{F}_{D_i}}=\mathbf{A V } (j2 \pi l T) \mathbf{E}[l] \mathbf{e}_i \mathbf{e}_i^T \mathbf{A}^T \mathbf{x}[l], ~ \forall i \in \mathcal{Q},
\end{eqnarray}
\begin{equation}
\frac{\partial \bm \eta[l] }{\partial \theta_{ai}}= \mathbf{\dot{A}} \mathbf{V} \mathbf{E}[l] \mathbf{e}_i \mathbf{e}_i^T \mathbf{A}^T \mathbf{x}[l] + \mathbf{A} \mathbf{V} \mathbf{E}[l] \mathbf{e}_i \mathbf{e}_i^T \mathbf{\dot{A}}^T \mathbf{x}[l] ,~\forall i \in \mathcal{Q},
\end{equation}
\begin{eqnarray}
\frac{\partial \bm \eta[l] }{\partial \bm\alpha_i }= \mathbf{A} \mathbf{V} \mathbf{E}[l] \mathbf{e}_i \mathbf{e}_i^T \mathbf{A}^T \mathbf{x}[l] [1, j],~\forall i \in \mathcal{Q},
\end{eqnarray}
where $\bm \alpha_i \triangleq [\Re\{\alpha_i\}, \Im\{\alpha_i\} ]^T$,  $\mathbf{e}_i$ is the $i$th column of $\mathbf{I}_Q$, and $\mathbf{\dot{A}} $ denotes the partial derivatives of $\mathbf{A} $ with respect to $\theta_{ai}$.
We note that
\begin{eqnarray}
\mathbf{F}_{\mathcal{F}_{D_i},\mathcal{F}_{D_j}} \!=\!2 \Re \Big\{ \mathrm{tr} \Big(\sum_{l\in \mathcal{L}} \frac{\partial^H \bm \eta[l] }{\partial \mathcal{F}_{D_i}} \mathbf{Z}^{-1} \frac{\partial \bm \eta[l] }{\partial \mathcal{F}_{D_j}} \Big) \Big\}, ~i,j \in \mathcal{Q }.\label{FIM_FF}
\end{eqnarray}
Since $\mathrm{tr}(\mathbf{AB})= \mathrm{tr}(\mathbf{BA})$, and $\mathbf{V}$ and $\mathbf{E}[l]$ are diagonal matrices, \eqref{FIM_FF} can be rewritten as
\begin{eqnarray}
\mathbf{F}_{\mathcal{F}_{D_i},\mathcal{F}_{D_j}}
& \!\!\!\!\!=\!\!\!\!\!& 2 \Re \Bigg\{ \mathrm{tr} \Bigg(
\sum_{l\in \mathcal{L}}
\big(\mathbf{A V } (j2 \pi l T) \mathbf{E}[l] \mathbf{e}_i \mathbf{e}_i^T \mathbf{A}^T \mathbf{x}[l] \big)^H
 \nonumber \\
&\!\!\!\!\!  \!\!\!\!\!& \times
  \mathbf{Z}^{-1} \big(\mathbf{A V } (j2 \pi l T) \mathbf{E}[l] \mathbf{e}_j \mathbf{e}_j^T \mathbf{A}^T \mathbf{x}[l]\big)
\Bigg)
\Bigg\} \nonumber \\
&\!\!\!\!\!=\!\!\!\!\!& 2\Re \Bigg\{ \mathrm{tr} \Bigg(
\sum_{l\in \mathcal{L}}
\mathbf{e}_i^T \mathbf{A}^H \mathbf{Z}^{-1} \mathbf{A} \mathbf{e}_j \mathbf{e}_j^T  \mathbf{V} \mathbf{E}[l](j2 \pi l T)
 \nonumber \\
&\!\!\!\!\!  \!\!\!\!\!&
 \times \mathbf{A}^T \mathbf{x}[l] \mathbf{x}^H[l] \mathbf{A}^* \mathbf{E}^H[l] (-j2 \pi l T) \mathbf{V}^H  \mathbf{e}_i
\Bigg)
\Bigg\} \nonumber \\
&\!\!\!\!\!=\!\!\!\!\!& 2\Re \Bigg\{ \mathrm{tr} \Bigg(
\sum_{l\in \mathcal{L}}
 (\mathbf{A}^H \mathbf{Z}^{-1} \mathbf{A})_{ij}  (\mathbf{V}^* \mathbf{A}^H \mathbf{R}_x^* \mathbf{A V})_{ij}
 \nonumber \\
&\!\!\!\!\!  \!\!\!\!\!& \times (\mathbf{\Sigma}_3)_{ij}
\Bigg)
\Bigg\}
,~i,j \in \mathcal{Q }.
\end{eqnarray}
where $(\cdot)_{ij}$ refer to the $i$th row and $j$th column element of the matrix. As a result, we obtain $\mathbf{F}_{\mathcal{F}_D\mathcal{F}_D} = 2 \Re\{\mathbf{F}_{44}\}$ with $\mathbf{F}_{44}$ specified in \eqref{F_44}. Similarly, other terms of $\mathbf{F}$ can be obtained in the same way by calculating the corresponding partial derivative, which results in the FIM in \eqref{FIM}.


\begin{balance}
\bibliography{IEEEabrv,IEEE_bib}
\end{balance}

\end{document}